\documentclass[12pt,manuscript]{aastex}
\usepackage{amsmath}
\usepackage{amsmath}
\usepackage{amsmath}
\usepackage{graphicx}
\usepackage{dcolumn}

\newcommand{\beq}{\begin{equation}}
\newcommand{\eeq}{\end{equation}}
\newcommand{\bea}{\begin{array}}
\newcommand{\eea}{\end{array}}\newcommand{\alp}{\alpha}

\begin{document}

\title{Modeling Planetary System Formation with N-Body Simulations:
 Role of Gas Disk and Statistics Comparing to Observations}
\author{Huigen Liu, Ji-Lin Zhou$^*$,  Su Wang}
\affil{Department of Astronomy \&  Key Laboratory of Modern Astronomy and Astrophysics in Ministry of Education, Nanjing University,
    China, 210093 \\
($^*$Email for correspondence: zhoujl@nju.edu.cn)}

\begin{abstract}
During the late stage of planet formation when  Mars-sized cores
appear, interactions among planetary cores can excite their orbital
eccentricities, accelerate  their mergings  and thus sculpture  their
final orbital architecture. This study contributes to the final
assembling of planetary systems with N-body simulations, including
the type I or II migrations of planets,  gas accretion of massive
cores in a viscous disk.   Statistics on the final distributions of
planetary masses, semimajor axes and eccentricities   are derived,
which are comparable to those of the observed systems. Our
simulations predict some new orbital signatures of planetary systems
around solar mass stars:  $36 \%$ of the survival planets are giant
planets ($>10M_\oplus$).  Most of the massive giant planets
($>30M_\oplus$) locate at 1-10AU. Terrestrial planets distribute
more or less evenly at $< 1-2$ AU.
  Planets in inner orbits may accumulate
at the inner edges of either the protostellar disk (3-5 days) or its
 MRI dead zone (30-50 days). There is a planet desert in the mass-eccecntricity
 diagram, i.e., lack of  planets with masses $0.005-0.08 M_J$ in highly eccentric orbits $(e>0.3-0.4)$.
 The  average eccentricity ($\sim 0.15$) of the giant planets  ($>10M_\oplus$) is
  bigger than that ($\sim 0.05$) of the terrestrial planets ($< 10M_\oplus$).
 A planetary system with more planets tends to have smaller planet masses
   and orbital eccentricities on average.

  \end{abstract}

\keywords{Methods: N-body simulations-
planetary systems: formation-planetary systems: protoplanetary  disk}

\section{Introduction}

 To date, around  490 exoplanets have been
detected\footnote{http://exoplanet.eu}, mostly by Doppler radial
velocity measurements(e.g., see Udry \& Santos 2007 for a review of
their statistics).  The masses of the planets  range from order of
Earth mass ($M_\oplus$) to tens of Jupiter masses ($M_J$). Among
them, the most recently observed  HD 10180 system records the most
numerous planets in exoplanetary systems(Lovis et al. 2010).  A
study of multi-planet systems is  helpful for understanding the
formation history of planetary architecture due to planetary
interactions (e.g., Wittenmyer et al. 2009). Recent statistics of
single and multiple planetary systems  have revealed several new
possible signatures(Wright et al. 2009):

\begin{enumerate}
  \item Including systems with long-term radial
velocity trends, at least 28\% of known  systems appear to contain
multiple planets. Thus multi-planet systems seem to be common.
  \item The distribution of  orbital distances
  of planets in multi-planet systems and single planets are inconsistent:
  single-planet systems show a pileup at period of $\sim 3$ days and a jump near 1 AU,
 while multi-planet systems show a more uniform
distribution in log-period.
  \item Planets in multi-planet systems have somewhat smaller eccentricities
  than single planets.
  \item Exoplanets with their minimum masses bigger
  than Jupiter  have eccentricities
broadly distributed across $0 < e < 0.5$, while lower mass exoplanets
exhibit a distribution peaked near $e = 0$.
  \item There may be a positive correlation
   between stellar masses and the occurrence rate of Jovian planets within 2.5AU (Johnson et al. 2007). This might be a reflection  of the
correlation of stellar masses with their circumstellar  disks.
\end{enumerate}

 Concurrently, our theoretical insights into the process of
planet formation have improved greatly.  According to the
conventional core accretion scenario, planets formed in
circumstellar disks. Through sedimentation of dust and cohesive
collisions of planetesimals, Mars-sized embryos will form through
runaway and oligarchic growth (Safronov 1969, Kokubo \& Ida 1998).
Angular momentum exchanges between the embryos and the gas disk will
cause a fast inward migration (type I) of the embryos in an
isothermal disk(Goldreich \& Tremaine 1979; Ward 1997; Tanaka et al.
2002).  In some circumstance, for example, in a region that
magnetorotation instability (MRI) is active (Laughlin et al., 2004,
Nelson \& Papaloizou 2004),  or in a radiative disk (Paardekooper \&
Mellema 2006, Kley et al. 2009), the speed of type I migration can
be greatly reduced or even with its direction being reversed. Thus
the embryos can be effectively sustained.
  As long as they grow beyond some critical masses ($\sim 10 M_\oplus $),
 quasi-hydrostatic gas sedimentation begins in the Kelvin-Helmholtz timescale
 (Pollack et al. 1996, Ikoma et al. 2000), which may last for million years
  until the final runaway gas accretion sets in.
   Type I migration of planetary cores is helpful to shorten the timescale of giant planet
formation, provided a factor of $\sim 10$ reducing of the migration speed (Alibert et al. 2005).
 The newly formed giant planets then undergo type II migration before the gas disk is
  depleted.

According to the above scenario of single planet formation,
planetary population synthesis  has reproduced successfully most of
the statistical characteristics for the observed systems. In a
serious of works,  Ida \& Lin (2004a, 2004b, 2005 and 2008)
predicted a planet desert with masses in 10-100 $M_\oplus$, due to
the runaway gas accretion of giant planets, and confirmed the
observed correlation that the planet detection rate increases with
the metallically of their host star(Fischer \& Valenti 2005). Other
studies with population synthesis of planets,  such as Kornet \&
Wolf (2006), Mordasini et al. (2009a,2009b), give more constrains of
planet formation model based on observational data.

However, planets formed in an environment with many siblings. Perturbations from
neighbor protoplanets may excite their  eccentricities,
 which is helpful to planet formation in some cases.
   For example, the inward type II
migration of gas giants in outside orbits may trap the planetary
cores  in inner mean motion resonances (mainly 2:1 resonance), thus
results in their mergings into hot super-Earths (Zhou et al. 2005,
Raymond et al. 2006).  While planetary perturbations may inhibit the
core growth in other cases, e.g., the lack of planets in the
location of asteroid belt may be due to Jupiter perturbations.
Moreover,  planet-planet scattering is thought as the major cause
for the eccentricities of the observed  planetary systems (Rasio \&
Ford 1996, Zhou et al. 2007, Chatterjee et al. 2008, Juri$\acute{c}$
\& Tremaine 2008). Planet population synthesis method without
including the planet-planet scattering procedure is incomplete to
account for the observed planetary signatures.

By taking the planet-planet scattering effect into consideration,
 N-body integration of the full planetary equations  is
necessary. Based on N-body simulations and a self-consistent disk
evolution,  Thommes et al. (2008) investigated the formation of
giant planets in the context of disk evolution, and revealed some
interesting tendencies that relate the planetary systems with their
birth disks. Ogihara \& Ida (2009) studied the formation and
distribution of terrestrial planets
 around M dwarf stars, and find the final configurations of  planets depend on
 the speed of type I migration.

 In this  work,
  we employ the N-body method to study the final assembly of planetary systems.
 The aims of this study are twofold: (1) to explain  the observed statistics
 of planet masses and orbit parameters, thus to constrain the parameters that are most suitable for planet formation;
 (2) to guide the future observation by extending our study
 to the mass regime that is not yet detectable.
  We adopt the standard model of core accretion scenario, which includes
 the standard type I and II migrations of planets.

 For N-body simulations of planetary formation and evolution, one
of the major difficulties is the initial conditions of planetary
embryos and the embedded protoplanetary disk.  At the later stage of
planet formation, most of the embryos are assumed to be formed with
their masses ranging from Mars to several Earths (or even bigger).
However, their radial distributions are quite uncertain. Also, the
parameters of viscous disks (their extensions, surface density
distributions, depletion timescales, etc.) are quite uncertain.
These parameters may be coupled, and their effects are different. As
shown in Thommes et al. (2008), disk properties play a key role in
determining the  planetary migration.  They chose the mass and the
viscosity of the protostellar disk as two free parameters.
 To simplify  the model used in this paper, through some tests,
we fix  most of the parameters to some fiducial  values,
except letting the gas disk depletion timescale ($\tau_{disk}$) and
disk mass ($\Sigma_{g}$)  as two  free parameters.
The effects of other parameters on the final architecture of the
planetary systems are also discussed.

 The arrangement of this paper is as follows. First we describe
 the model and method of the paper in \S 2,. Then   some analytical
 estimation about the orbital configuration under type I and II migrations
 is given in \S 3.  The  effects of disk parameters on the evolution and final
 architecture of the planetary systems  are  discussed in \S 4.
 In \S 5 , the distributions of planet parameters (masses, semimajor axes, eccentricities)
 from simulations are compared with observations. To see if our results are credible in two-dimensional model , we survey
 the differences in three-dimensional model in \S 6.
 The conclusions are presented in \S 7, with discussions and some implications for future works.

\section{Model and Initial Setup}

Our investigation  starts  at the stage that most of the
planetary embryos have cleared up their feeding zones and
obtained their isolation masses. This may correspond to an epoch of
 $\sim 1$ Myr after the birth of the protostar.  Embryos in inner obits might
 undergo type I migration, which is assumed to be
 stalled at some locations such as the
inner edge of MRI dead zone of the gas disk (Kretke \& Lin
2007, Kretke et al. 2009). For embryos outside the snow line,
some of  them  are large enough (several $M_\oplus$) for the onset
of efficient gas accretion. They begin to accrete  gas, open gaps, and undergo
type II migration. Detailed model and parameters that we used
in this paper are presented as follows.

\subsection{Viscous Disks}

Pre-main-sequence stars accrete gas through circumstellar disks.
Although the origin of the disk viscosity is still in controversial,
 the onset of MRI helps to the transportation of angular momentum through the
disk (Balbus \& Hawley 1991).
In this paper, we adopt the {\it ad hoc} $\alpha$-prescription (Shakura \& Sunyaev 1973)
 to model the effect  of mass transportation during the
 stage of classical T-Tauri stars.
In such a disk, the kinematical viscosity is expressed as $\nu=\alpha
c_s h$, where $c_s$ and $h$ are the sound speed in mid-plane disk and the
density scale height, respectively(Table 1).  The surface density at
stellar distance $a$ is given as (e.g., Pringle 1981),
 \beq
   \Sigma_g=\frac{\dot{M}}{3\pi \alpha c_s h
   }\left[1-\left(\frac{R_*}{a}\right)^{1/2}\right],
   \label{sigg1}
\eeq where $\dot{M}$ is the gas accretion rate and
 $R_*$ is the stellar radius.  For a T-Tauri disk with stellar  mass of
$0.2M_\odot \le M_* <2.0M_\odot$, the observed accretion rate is
(Natta et al. 2006,  Vorobyov \& Basu 2009) \beq
  \dot{M}= 2.5  \times 10^{-8} M_\odot
   {\rm yr}^{-1} \left(\frac{M_*}{M_\odot}\right)^{1.3\pm 0.3}.
\eeq In an optically thin disk,  adopting the parameters in Table 1
and let $a \gg R_*$, we get Eq.(\ref{sigg2}) from equation
(\ref{sigg1}),
 \beq
 \Sigma_g=2400 {\rm ~g~cm^{-2}}
 \left( \frac{\alpha}{10^{-3}}\right)^{-1} \left(\frac{M_*}{M_\odot} \right)^{0.8}
  \left( \frac{a}{1\rm ~AU} \right)^{-1}.
\label{sigg2}
 \eeq

 For an $\alpha$-disk  with constant $\alpha = 10^{-3}$, we  derive the
 minimum mass solar nebular (MMSN)  except with a
  different slope, $\beta=-\ln \Sigma_g/\ln a$ (c.f., $\beta= 3/2$ in MMSN, Hayashi 1981,Ida \& Lin 2004a).
We also assume that the protoplanetary disk has a layered structure as a sandwich:
inside  a location $a_{\rm crit}$,  the protostellar disk is thermally ionized partly,
 while outside  $a_{\rm crit}$ only the
 surface layer is ionized by stellar X-rays and diffuse cosmic rays, leaving the
central part of the disk a highly neutral and inactive
``deadzone"(Gammie 1996). The viscosity in the
MRI active region could be one or two magnitudes  larger than  that of the
 dead zone(Sano et al. 2000). By assuming a constant accretion rate across the
disk at a specific epoch in equation (1),  a positive density gradient is expected
near $a_{\rm crit}$,  which helps to halt the embryos
 under type I migration(Kretke \& Lin, 2007, Kretke et al. 2009).

To model this effect, we let $\alpha_{\rm MRI}$ and $\alpha_{\rm
dead}$ denote the $\alpha$-values of the MRI active and dead
regions, respectively. The effective $\alpha$ for the disk is
modeled as (Kretke \& Lin, 2007):
\begin{equation}
\alpha_{\rm eff} (a)=\frac{\alpha_{\rm dead}-\alpha_{\rm MRI}}{2}
[{\rm erf}(\frac{a-a_{\rm crit}}{0.1a_{\rm crit}})+1]+\alpha_{\rm
MRI},
 \label{alpeff}
\end{equation}
where ${\rm erf}$ is the error function,  $0.1 a_{\rm crit}$ is
thought as the width of the transition region. In this paper, we
adopt $\alpha_{\rm MRI}=0.02,\alpha_{\rm dead}=10^{-4}$ (Sano et al.
2000).
 The small value of $\alpha_{\rm dead}$ is adopted to obtain a reasonable
 timescale of type II migration (see Eq.[{\ref{migII}] later).    Alternative choices
 of $\alpha_{\rm dead}$ are also discussed in \S4.1.

The location of the inner edge of the MRI dead zone, $a_{\rm crit}$,
 varies with the disk temperature, kinematics and
mass accretion rate, etc. Here we adopt the expression from Kretke et al. (2009),
\begin{equation}
a_{\rm crit}=0.16 ~{\rm AU}(\frac{\dot{M}}{10^{-8}M_{\odot}{\rm yr}^{-1}})^{4/9}
(\frac{M_{*}}{M_{\odot}})^{1/3}(\frac{\alpha_{\rm MRI}}{0.02})^{-1/5}.
 \label{acrit}
\end{equation}
During the evolution of a T-Tauri star and its disk, $\dot{M}$
decreases from $\le 10^{-6}$  to $\sim 10^{-9}$ $~{\rm M}_\odot$
yr$^{-1}$, according to the infrared  excess observation(e.g.
Gullbring et al. 1998). So the location of $a_{\rm crit}$ is
migrating inward as  time proceeds. To simplify the procedure, we
let the gas disk depletes uniformly in a timescale $\tau_{\rm disk}$,
so $\dot{M}=\dot{M}_0\exp(-t/\tau_{\rm disk})$. Assuming a  disk
mass of $0.02M_{\odot}$ (Andrews \& Williams 2005), we obtain
$\dot{M}_0=0.02M_{\odot}/\tau_{\rm disk}$.

Substituting  the previous equations (\ref{alpeff}) and (\ref{acrit}) into
 equation (\ref{sigg2}),   we obtain the following surface density for the circumstellar disk:
\begin{equation}
\Sigma_{g}=\Sigma_{0} f _g   (\frac{a}{\rm
1AU})^{-1}(\frac{\alpha_{\rm eff}}{10^{-4}})^{-1}(\frac{M_{*}}{M_{\odot}})^{4/5}
\exp({-\frac{t}{\tau_{\rm disk}}}),
\label{sigg3}
\end{equation}
where $f_g$ is the gas enhancement factor, $\Sigma_{0}=280 $g
cm$^{-2}$ is  adopted in this work so that the total mass of the
disk up to 100 AU is $0.02M_{\bigodot}$ for $f_g=1$, corresponding
to the average disk mass in Taurus-Auriga star formation
region(Beckwith \& Sargent 1996).
 For such a disk, the Toomre's criterion for the onset of
 gravitational instability is expressed as $Q=\frac{c_s\Omega_K}{\pi G\Sigma_g}
 =340 (a/1AU)^{-3/4}f_g^{-1}$. So $Q>1$ holds  and gravitational
 instability will not occur  up to $100$ AU as long as $f_g \le 10$.

 We truncate the  inner disk at the disk cavity due to the stellar magnetic field.
Around the corotation radius, the stellar magnetic torque acts to
extract angular momentum from the disk and spins down disk
material. 
At the location where the stellar magnetic field completely
dominates the disk internal stresses, sub-Keplerian rotation leads
to a free-fall of disk material onto the surface of the star
 in a funnel flow along magnetic-field lines, which results in an inner disk
 truncation\cite{Kon91}.
 The maximum distance of the disk truncation is estimated at $\sim 9$
 stellar radii. Considering  that the radius of protostar is generally 2-3 times
  larger than their counterparts in the main sequence stage,
  the inner disk truncation would occur at $<  0.1$AU.
Thus we set the inner edge as 0.05AU in this paper.

\subsection{Isolated Embryos and Type I Migration}

According to the core accretion scenario of planet formation,
planetary embryos form through cohesive  collisions of heavy
elements near the midplane of the circumstellar disk. Within the
solid disk  an embryo  will
grow 
until it accretes all the dust material round its feeding zone
so that an isolation body  is achieved (Kokubo \& Ida 1998,Kokubo \& Ida 2002).
 At the later  stage of planet formation,  inward migration of
planetesimals under gas drag and planetary embryos under type I
migration may change the distribution of embryos, which results in
an uncertainty of the initial embryos masses. On the other hand, the
MMSN model for the solid disk was derived according to the final
solid masses of the present planets in  the solar system (Hayashi
1981, Ida\& Lin 2004a),  which reflects the final distribution of
the embryo, so we adopt the isolation masses with distribution as
that of MMSN but with an enhancement factor $f_d=f_g$. In a gas
disk, the instability of the isolation masses is suppressed until
the gas disk is depleted. Assume the dynamical instability timescale
of a group of isolated masses is the same as  the disk dispersal
timescale, the isolation masses  are correlated with their mutual
separations, and are given by (Kokubo \& Ida 2002, Zhou et al. 2007)
\begin{equation}
M_{iso}=0.16 M_{\oplus} ~ (f_d \eta_{\rm ice})^{3/2}(\frac{a}{1AU})^{3/4}(\frac{M_*}{M_{\bigodot}})^{-1/2} \left(\frac{k_{\rm iso}}{10} \right)^{3/2},
\label{msio}
\end{equation}
where $f_d$ is the heavy elements enhancement  factor over MMSN model,
$\eta_{\rm ice}=1$ inside $a_{\rm ice}$ and $4.2$ outside $a_{\rm ice}$,  with
$a_{\rm ice}$  the location of the snow line,
beyond which water is condensed as ice from disk gas ($T \simeq 170K$).
$k_{\rm iso}$ is the  separation of embryos scaled by $R_H=(2M_{\rm iso}/3M_*)^{1/3}a$.
For $\tau_{\rm disk}=1-10$ Myr, $k_{\rm iso}=8-10$ at 1AU and $7-9$ at 10 AU
for a disk with $f_d=1$ (See Fig.11 of Zhou et al. 2007 and details therein).


 Due to the variation of the disk accretion rate  and
 stellar radiation at different epoch of the  protostar,
the location of snow line varies (Garaud \& Lin 2007).
For a star with mass $\le  3 ~M_\odot$,
 we adopt the snow line location as Kennedy \& Kenyon (2008):
\begin{equation}
a_{ice}=2.7~{\rm AU}(\frac{M_*}{M_{\odot}})^{4/9}(\frac{\dot{M}}{10^{-8}M_{\odot}~{\rm yr}^{-1}})^{2/9}.
\label{aice}
\end{equation}

To model the type I migration of the embryos, we adopt the
expression of the migration timescale from Cresswell
\& Nelson (2006), which is also valid for  eccentric  orbits,
\begin{eqnarray}
\tau_{\rm mig,I}&=&-\frac{1}{C_1}\frac{1}{2.7+1.1\beta}
\left(\frac{M_*}{M_p}\right)\left(\frac{M_*}{\Sigma_g a^2}\right)
 \left( \frac{h}{r}\right)^2\nonumber \\
& &\left| \frac{1+(\frac{er}{1.3h})^5}{1-(\frac{er}{1.1h})^4}
\right|\Omega^{-1}, \label{migI}
\end{eqnarray}
where negative (positive) value of $\tau_{\rm mig,I}$ corresponds to
the inward(outward) migration respectively, $M_p$, $r$, $e$,
$\Omega$ are the  mass, position,  eccentricity, angular velocity of
the planet, respectively. Due to the MRI effect, $\beta \equiv
-\partial {\rm ln} \Sigma_g/ \partial {\rm ln} a$ can be negative
near the location of maximum pressure $a_{\rm crit}$. $C_1$ is a
reduction  factor.   Lots of literatures have shown that, to produce
the observed planetary occurrence rate, $C_1 \in [0.03,0.3]$ is an
appropriate  range (e.g. Alibert et al. 2005, Ida \& Lin 2008). To
test its validity in our model,  we set $C_1=0.03,0.1,0.3$ and
execute some test runs. As shown in Fig.1a, both the choice of
$C_1=0.03$ and $0.1 $ produce a very slow planet migration embedded
in our disk model,
 which may result in  the formation of hot Jupiters very difficult.  So we set $C_1=0.3$ throughout this
 paper.

The presence of the gas disk will damp the eccentricities of
embedded embryos(Goldreich \& Tremaine 1980). The  $e$-damping
timescale can be described as (Cresswell \& Nelson 2006),
\begin{equation}
\tau_{\rm edap,I}=\frac{Q_e}{0.78}\left(\frac{M_*}{M_p}\right)
\left(\frac{M_*}{a^2\Sigma_g}\right)\left(\frac{h}{r}\right)^4
\left[1+\frac{1}{4}(e\frac{r}{h})^3\right]\Omega^{-1}, \label{edpI}
\end{equation}
where 
 $Q_e=0.1$ is a normalization factor to fit with the hydrodynamical simulations.

\subsection{Giant planet formation: gap opening and type II migration}

According to the conventional   accretion
scenario, giant planets form through three major
stages(Perri \& Cameron 1974; Mizuno 1980; Pollack et al. 1996) : (1) Embryo growth stage. Protoplanetary cores form and
grow mainly by the bombardment of planetesimals before they attain  isolation masses.  (2) Quasi-hydrostatic sedimentation stage. The accretion of planetesimals tapers as their supply in the feeding zone is depleted.  This induces a quasi-hydrostatic sedimentation and the growth of the gaseous envelope due to the loss of entropy. (3) Runaway gas-accretion stage. When the mass of gas envelop
becomes comparable to that of the core, a runaway stage
of gas accretion sets in  continually until the gas supply is
exhausted by either the formation of a tidally induced gap near
the protoplanet orbit or the depletion of the entire  disk.

Through quasi-static evolutionary simulations, Ikoma et al. (2000)
derived  the critical core mass where significant gas accretion
occurs: \beq M_{\rm crit} \sim  7 M_\oplus \left(\frac{\dot M_{\rm
core}}{10^{-7} M_\oplus {\rm yr}}\right)^{0.25}
\left(\frac{\kappa}{1 ~{\rm cm^2~ g}^{-1}}\right)^{0.25},
\label{mcrit} \eeq where $\dot M_{\rm core}$ is the rate at which
planetesimals are accreted onto the core,  and $\kappa$ is the grain
opacity (see also Rafikov 2006).   Due to the uncertainty  of $\dot
M_{\rm core}$ and $\kappa$, we assume $\dot M_{\rm core} \propto
f_g$ and adopt
  $M_{\rm crit} = 4M_{\oplus} f_g^{0.25}$  in this paper.
A planet core beyond this critical mass  will  begin
to  accrete gas in a Kelvin-Helmholtz  timescale, $\tau_{\rm KH}\simeq
10^9 {\rm~yr} (M_p/M_\oplus)^{-3}$, so that
$dM_p/dt\simeq M_p/\tau_{\rm KH}$(Ikoma et al.  2000).
However, the accretion rate is also limited by
the replenishing rate of the materials ($\dot{M}_{\rm disk}$), so
the accretion rate of the planetary embryos is expressed as(Ida \& Lin  2004),
\begin{equation}
\frac{dM_p}{dt}=\min \left[ 10^{-9}(\frac{M_p}{M_{\oplus}})
^4M_{\oplus}{\rm yr}^{-1},\dot{M}_{\rm disk} \right] \label{dmp}.
\end{equation}

When the planet mass grows to sufficiently large, a tidal-induced gap in
the gas disk forms around its orbit (Lin \& Papaloizou 1979).
The critical mass for  gap-opening
is determined  by equating the timescale for Type I torques to open a gap
 (in the absence of viscosity) with
that for viscous diffusion to fill it in.  This gives
(Armitage \& Rice, 2005)
\beq
\frac{M_p}{M_*}\geq\alpha_{\rm eff}^{1/2}(\frac{h}{a})^2 \label{mgap}.
\eeq
With $\alpha_{\rm eff} \sim 10^{-4}$ in the MRI dead zone,
the critical mass is given by:
\begin{equation}
M_{I,II}=7.5(\frac{a}{\rm 1~AU})^{1/2}(\frac{M_*}{M_{\bigodot}})M_{\oplus}.
\label{MIII}
\end{equation}

After the giant planet opens a gap around the disk,
it is embedded in the viscous  disk and undergos  type II migration.
 As the mass of the planet grows and becomes comparable
to the disk mass, migration slows down and eventually stops.
The migration speed with the slow down effect can be
expressed as(Alibert et al. 2005):
\begin{equation}
\frac{da}{dt}=-\frac{3\nu}{2a}\times \min(1,\frac{2\Sigma_ga^2}{M_p}).
\label{dapdt}
\end{equation}
So the migration timescale is given as:
\begin{eqnarray}
\tau_{\rm mig,II}&=&0.6 {\rm ~Myr}~(\frac{\alpha_{\rm
eff}}{10^{-4}})^{-1} (\frac{a}{\rm
1AU})(\frac{M_*}{M_{\odot}})^{-1/2}\nonumber\\
& &\ max(1,\frac{M_p}{2\Sigma_ga^2}). \label{migII}
\end{eqnarray}

During the migration of the giant planets, their eccentricities
will be damped by the disk tide(Goldreich \&
Tremaine, 1979, 1980; Ward 1988), unless  the
 planet is very massive ($\sim 20~ M_J $, Papaloizou et al. 2001).
 As the $e$-damping rate  due to the gas disk is  quite elusive
 for different mass regimes  of  the
 planets, we adopt an empirical formula
for the $e$-damping timescale (Lee \& Peale 2002) \beq \tau_{\rm
edap,II}= \tau_{\rm mig,II}/{\rm K},
 \label{edpII} \eeq
where $K$ is a positive constant with a value ranging $10-100$.  To
choose an appropriate value of $K$, we  execute some test runs with
$K=10,30,100$. As shown in Fig.1b, the planet eccentricity is damped
very quickly with $K=100$.  For $K=30$, the eccentricity can be
excited and  effectively damped  at the end of  simulation when the
gas disk has been nearly depleted. Therefor we take $K=10$ so that
some planets in eccentric orbits can finally survive according to
the observations.

The gas accretion of a planet will stop when all the gas material around the
feeding zone($R_{H}=h$) is exhausted, this gives the truncation mass  of a gas giant:
\begin{equation}
M_{\rm g,iso}(a)  =
120(\frac{a}{1AU})^{3/4}(\frac{M_*}{M_{\odot}})M_{\oplus}.
\label{mtran}
\end{equation}
This  limits the  mass of a giant planet by pure gas accretion.
However,  collision and cohesive mergings may increase the masses of
giant
 planets up to several Jupiter masses.

\subsection{Equation of Motion}

Based on the model described above, we investigate the late stage
formation  of the planetary system in this paper. The stellar mass
is taken as $1~M_\odot$. The gas disk is assumed as in Equations
(\ref{sigg3}). We further assume that embryos have obtained their
isolated masses (see Eq.[\ref{msio}]) with $f_d=f_g$.
 We ignore
those embryos with mass $\le 0.1~M_\oplus$ in the  inner and outer
orbits, so that there are totally $38\sim44$ isolated embryos
initially in circular and coplanar orbits in [0.5-13.5]AU for  a
system with a gas  and a solid disk  $f_g=f_d=1$. The isolation
masses and their radial extension will be changed accordingly for
different $f_g=f_d$.
The outer boundary of embryos is set according to the rule that,  beyond which the core masses that can grow within 5 Myrs are less than 0.1 $M_\oplus$, according to the standard core growth (Kokubo \& Ida 2002, Ida \& Lin 2004a).
 Although this simplification may  neglect  their dynamical frictions to  inner massive cores, this effect is similar to the damping effect by the gas disk tide,  which is more effective  and included already in our simulations.

The angle elements (longitude of periastron, mean motion) of the embryos
  are randomly chosen.  Fig.2 shows  an example of the masses and initial locations of the embryos.
  According to equation (\ref{msio}), different $\tau_{\rm disk}$ gives slightly  different values  of $k_{\rm iso}$ and $M_{\rm iso}$(see Zhou et al. 2007).

 The acceleration of an embryo  with mass $M_i ~(i=1,...N)$  is given as,
\begin{eqnarray}
\frac{d}{dt}\textbf{v}_i&=& -\frac{G(M_*+M_i)\textbf{r}_i}{r_i^3} +
\sum _{j\neq i}^N GM_j
\left[\frac{\textbf{r}_j-\textbf{r}_i}{|\textbf{r}_j-\textbf{r}_i|^3
} - \frac{\textbf{r}_j}{r_j^3}\right]\nonumber\\
& &\ -\frac{{\bf v}_i}{2 \tau_{\rm mig}}-\frac{(\textbf{v}_i \cdot
\textbf{r}_i)\textbf{r}_i}{r_i^2\tau_{edap}}, \label{eqf}
\end{eqnarray}
where $\textbf{r}_i$ and $\textbf{v}_i$ are the position and
velocity vectors of $M_i$ relative to the star, the third and fourth
terms in the r.h.s. of the equation are the accelerations that cause
the eccentricity damping and migration, respectively. For embryos
with $M_i< M_{I,II}$ defined in Eq.(\ref{MIII}), $\tau_{\rm
mig}=\tau_{\rm mig,I}$ (Eq.[\ref{migI}]) and $\tau_{\rm
edap}=\tau_{\rm edap,I}$ (Eq.[\ref{edpI}])
 are used.
For those with  $M_i>  M_{I,II}$, $\tau_{\rm mig}=\tau_{\rm mig,II}$ (Eq.[\ref{migII}])
and $\tau_{\rm  edap}=\tau_{\rm edap,II}$ (Eq.[\ref{edpII}])  are used instead.
Note that, during the growth of an embryo, it may cross the critical mass
$M_{I,II}$, thus pass from  type I to type II migration, and also
the eccentricity-dampping  mode is switched.

In the absence of mutual planet perturbations ($M_j=0, ~j\neq i)$, i.e., without second term
in the r.h.s. of Eq.(\ref{eqf}),
the secular variations of orbital elements for the embryo $M_i$
under migration and eccentricity-damping are derived from the
classical perturbation theory:
\beq
\begin{array}{l}
\left< \frac{da}{dt}\right> =-\frac{a}{\tau_{\rm mig}}-\frac{2e^2}{\tau_{\rm edap}(1+\sqrt{1-e^2})}, \\
\left<\frac{de}{dt}\right> =-\frac{e(1-e^2)}{\tau_{\rm edap}(1+\sqrt{1-e^2})}, \\
\left<\frac{d\omega}{dt}\right>=0
\end{array}
\eeq where $\omega$ is the argument of periastron of the embryo
orbit.

\section{Planet Configurations Under Migrations:  Analytical Considerations}

 Before we present the numerical results, it will be useful to investigate  some
 ideal cases that  some cores undergo type I or II migrations only,
 and to see the configuration  of the system without considering their mutual perturbations.
This will be helpful to  understand the onset of
 instability for the full system.

 \subsection{Planetary core configurations under type I migration}

Suppose there are N planet cores with masses below $M_{I,II}$ so that they
undergo type I migration. Eq.({\ref{migI}) gives
 $\tilde{a}/\dot{\tilde{a}}=\tau_{\rm mig,I}= k \tilde{M}_p^{-1}  \tilde{a} \exp (t/\tau_{disk})$ for
$a \gg a_{\rm crit}$,
 where  $\tilde{M}_p=M_p/M_\oplus$, $\tilde{a}=a/{\rm 1~AU}$ and $k \approx 0.23$Myr if $C_1=0.3$ with the expression of $\Sigma_g$ from Equation (\ref{sigg3}).
An embryo with initial semimajor axis $\tilde{a}_0$
will evolve with
$ \tilde{a}=\tilde{a}_0- \xi \tilde{M}_p$,
where $\xi=\tau_{\rm disk}[1-\exp(-t/\tau_{\rm disk})]/k$.
 So the evolution of relative separation  between  two neighboring
embryos with mass difference $\Delta \tilde{M}_p$ is given as,
\beq
 \frac{\Delta \tilde{a}}{\tilde{a}}
 =\left(\frac{\Delta \tilde{a}_0}{\tilde{a_0}}\right) \frac{1-\xi \Delta \tilde{M_p}/\Delta \tilde{a}_0}{
 1-\xi \tilde{M}_p/\tilde{a}_0},
 \label{daa1}
\eeq
where $\Delta \tilde{a}_0$ is the initial separation.
For a general case, $\Delta \tilde{M}/\Delta \tilde{a}_0 = \gamma \tilde{M}/ \tilde{a}_0$,
and  since $ \xi \tilde{M}_p/\tilde{a}_0 < 1$ before $a$ goes to $a_{\rm crit}$, Eq.(\ref{daa1})
approximates to,
\beq
 \frac{\Delta \tilde{a}}{\tilde{a}}
 \approx \left(\frac{\Delta \tilde{a}_0}{\tilde{a_0}}\right) [1+(1-\gamma) \xi \tilde{M}/\tilde{a}_0 ].
 \label{daa2}
\eeq So, as long as $\gamma<1$ , the inward type I migration will
increase the mutual distance scaled by their mutual Hill radii
($R_H\propto a$ ). According to Zhou et al. 2007, such an increase
will enhance
 the orbital crossing timescale of the system, for example,
 in the case of isolation masses in Eq.(\ref{msio}) with $\gamma=3/4$.
However, due to the perturbation of giant planets and their mutual
perturbations, embryos in inner orbits will have their
eccentricities excited, which may result in
 instability.

 \subsection{Giant planet configurations under type II migration}

Similarly  to the previous analysis,
giant planets undergoing type II migration with
a timescale of Eq.(\ref{migII}) may modify  their mutual separations.
With the same notion of previous subsection,
we assume that  the planet mass is much smaller than the inner disk mass,  and the disk
has a constant $\alpha_{\rm eff}$. Thus  one has
 $\tilde{a}/\dot{\tilde{a}}=\tau_{\rm mig,II}=
  k'  \tilde{a} $ for $a \gg a_{\rm crit}$, where
  $k'\approx  0.6$ Myr from equation (\ref{migII}).
  So $\tilde{a}=\tilde{a}_0-t/k'$, and two neighboring planets  will have
constant separation $\Delta \tilde{a}_0$. When it is
 scaled by $R_H\propto  a$ at time $t$,
\beq
 \frac{\Delta \tilde{a}}{\tilde{a}}
 =\left(\frac{\Delta \tilde{a}_0}{\tilde{a_0}}\right)
 \left( 1-\frac{t}{k'\tilde{a}_0 } \right)^{-1},  ~~(t<k'\tilde{a}_0).
 \label{daa}
\eeq
This means that their mutual separation scaled by Hill radii  will
increase during the inward type II migration. However, secular perturbations among them
will excite their eccentricities during their convergent migration, thus
destabilize the system if their eccentricities are high enough.

\section{Numerical Results: Dependence on Disk Parameters}

We numerically integrate the equations of planet motion (\ref{eqf})
with a time-symmetric Hermite scheme (Kokubo et al. 1998, Aarseth
2003). Regularization technique is used to handle the collisions
between embryos,  all the embryos have their physical radii (a mean
density of $3$ g cm$^{-3}$ is assumed), and mergings are expected
when the mutual distance of two embryos  is less than the sum of
their physical radii. We assume a perfect inelastic collision
between embryos. The inner boundary of the gas disk is set as
$0.05$AU. If a planet migrates to $<0.04$AU,  we remove it from
further integration. The external boundary of the gas disk is set as
$100$AU. If a planet evolves to the orbit  with  $a>50$AU, we
consider it as   being escaped from this system.  The tidal effects
between the host star and the close-in planets are not included in
our model for their long effective timescales.

Figure 2 shows an example of such an evolution.
 Initially 44 embryos are put, with their masses and initial locations
 shown in Fig.2(a).  Finally, three planets are left, with masses
  $3.1 M_\oplus$, $238.2 M_\oplus$ and  $137.6 M_\oplus$
 (from inner to outer).  Moreover,  the outer two planets passed  the periastron alignment (with $\varpi_2-\varpi_3$ librating around 0) during the evolution.

 As we mentioned,  parameters of the protoplanetary disk are quite
uncertain with the present ability  of observations. So in this
section, we mainly investigate the effect of the disk mass,
viscosity and depletion timescale on the planet evolution.

\subsection{Influences of the disk viscosity ($\alpha_{dead}$) and disk mass ($f_g$)}

The variations of the disk mass and viscosity are modeled by two
parameters, i.e.,  $f_g$ in Eq.  (\ref{sigg3})  and $\alpha_{\rm
dead}$ in Eq.(\ref{alpeff}). We set $f_g=0.3,1,3,10$
 and  $\alpha_{dead} =10^{-2},10^{-3},10^{-4}$ to
survey their contributions to the architecture of planetary systems
individually. Five runs for each parameter are executed.
We fix the surface density profile as well as the disk depletion timescale ($\tau_{\rm disk}=2{\rm Myr}$) for these test runs.

In our model $\alpha_{\rm dead}$ mainly works on type I migration
via Eq.(\ref{migI})  and type II migration via Eq.(\ref{migII}).
Since $\Sigma_g \propto (\alp_{eff})^{-1}$ (Eq.[\ref{sigg3}]), thus
$\tau_{\rm mig,I}\propto \alp_{eff}$,  i.e.,  planets in disks with
smaller $\alpha_{eff}(\sim 10^{-4})$ migrate faster, which results
in a smaller average semimajor axis(Fig.3a). For type II migration,
if the planet mass is small ($M_p\le 2\Sigma_g a^2$, which is about
$90M_\oplus $ at 5AU and $170M_\oplus$ at 10AU), the migration speed
decreases with $\alpha_{eff}$, thus larger $\alpha_{eff}$ ($\sim
10^{-2}$) may also lead to a  small  average semimajor axis. Thus
the disk viscosity may change the final location of the planets. So
we take a most plausible value ($\alpha_{\rm dead}=10^{-4} $) in the
following simulations(Sano et al. 2000).
 On the other hand,  the average eccentricities  are around 0.1 with large variations  for all cases, which
seems to be insensitive with $\alpha_{\rm dead}$.
  The reason is that   small $\alpha_{\rm dead}$ leads to a high density of gas
disk,  resulting  in two competing effects: (a)  a fast type I migrations so that  a large frequency of embryo-scatterings is
expected, which enhance the eccentricities of embryos. (b) a fast $e$-damping rate with   a small $e_{\rm mean}$.  As the timescales
of these two effects are comparable,   $e_{\rm mean}$ is nearly independent of $\alpha_{\rm dead}$.

There are three ways for $f_g$ to influence the orbital evolutions:
the surface density  of the gas disk(Eq.[\ref{sigg3}]) ,   initial
isolation masses (Eq.[\ref{msio}] with $f_d=f_g$), and the
  critical mass for the onset of runaway gas accretion (Eq.[\ref{mcrit}]).
Larger $f_g$ may result in larger  cores extended to an outer
region. As we assume $\dot{M}_{\rm core} \propto f_d$, we have
$M_{\rm crit}\propto f_g^{0.25}$. Comparing with the masses of the
initial embryos($\propto f_g^{3/2}$), it's much easier to form
massive gas giants  in a massive gas disk.

As shown in Fig.3b, when $f_g\geq1$, $a_{\rm mean}$ are all around
$3-4$AU. Although the migrations should be faster for planets in
disks with larger $f_g$, the initial embryos extended to an outer
region in these cases, which  results in similar averaged locations.
For a less massive disk($f_g=0.3$), these  small embryos  initially
in the inner region have never accreted gas, and  thus they
experienced   type I migration throughout the evolution, therefore
they have smaller $a_{\rm mean}$ and $e_{\rm mean}$. More effective
eccentricity-dampping  results in a smaller $e_{\rm mean}$ for
$f_g=3$ than that of $f_g=1$. When $f_g=10$,  the largest embryos
can reach  $~80M_{\oplus}$ via collisions in 1Myr.  Finally the
planets left in these systems are mainly gas giants, and their
eccentricities are damped  much less effectively than those  of
small planets.  This is why $e_{\rm mean}$ of $f_g=10$ is much
larger than that of $f_g=3$.

\subsection{Correlations with the disk lifetime ( $\tau_{\rm disk}$)}

 The role of the gas disk on the planet evolution  is manifold:
it causes the inward migration (type I or  II)  of protoplanets 
 as well as the tidal damping of their orbital eccentricities. 
A long survival gas disk may be helpful to the formation of giant
planets, while  planets in a short-lived disk may not have enough
time to accrete gas,  thus remain either terrestrial or Neptunian
planets. To  investigate the effect of the disk depletion timescales
($\tau_{\rm disk}$), we set 11 values of $\tau_{\rm disk}$, evenly
ranged from 0.5 Myr to 5Myr in a logarithm scale (Haisch et al.
2001).
 For each $\tau_{\rm disk}$ we did  20 runs of simulations by choosing the orbital phase angles of the
embryos randomly. So totally we did 220 simulations for $f_g=1$. All
the simulations are stopped at $t=10$ Myr.

An  interesting problem for
planet formation is the growth epoch of different types of
planets. We define (somewhat arbitrary) the  following
two types of planets:
 gas giant planets (GPs), includes massive GPs ($M_p \ge 30M_\oplus$)
 and Neptune-sized GPs ($10M_\oplus \le M_p<30M_\oplus$);
 terrestrial planets (TPs), including Super-Earth TPs ($1M_\oplus \le M_p<10M_\oplus$)
 and Sub-Earth TPs($M_p<1M_\oplus$).
 Fig.4 shows the distribution of planet semimajor axes for the 220 runs of simulations
 at some epoches with $f_g=1$.  Only TPs are present at $t=10^4$yr (Fig.4a).
  When $t=1$Myr, runway gas accretion occurred, thus a few GPs appear (Fig.4b).
  The most efficient growth of GPs occurred at $1-3$Myr,
  thus the number of GPs increased very fast (Fig.4c), until they become
   the dominant members at the end of simulations(Fig.4d).

Fig.5 plots the correlations between the properties  of the final
systems and $\tau_{\rm disk}$.
Basically there are two regimes. 
 For short-lived disks with $\tau_{disk} \le 1$Myr,  the forming
planets have small average masses ($70-200 ~M_\oplus$, Fig.5c), the
average semimjor axes of these systems are large  ($5-6 $AU,
Fig.5d), indicating most of the  surviving planets were formed  in
distant orbits, while  migrations have not affected their orbital
architectures strongly.  The average eccentricities ($\sim 0.15$)
are relatively big (Fig.5a), showing the tidal damping effect is not
very effective due to the short lifetime of the disk. On the
contrary,  planetary systems with a longer disk lifetime
($\tau_{disk}>1 $Myr) tend to have larger average planet masses
($\sim 1$ Jupiter mass ), with lower averaged eccentricities ($\sim
0.1$) due to the long period of the disk damping. The semimajor axes
of these systems decrease  from   4 AU to 2 AU, indicating inward
migration  indeed plays an  important role in sculpting their
orbital configurations.

 Systems with $\tau_{\rm disk}<1.5$Myr do not have much differences on the surviving number of the planetary systems,  while $N_{\rm
mean}$ decrease slightly  as $\tau_{\rm disk}>1.5$Myr increasing (Fig.5b). The maximum of $N_{\rm mean}$ at $\sim 1.5$Myr, is due to
the mechanism of halting small planets near $a_{\rm crit}$ (see \S 5.1). In disks with smaller $\tau_{\rm disk}$, the effects due to
gas disks are not so efficient, and the surviving number of planet systems mainly depends on the interactions between embryos. In the
disks with longer lifetime,  giant planets experience sufficient type II migration and will be closer to the boundary of MRI region,
where small planets  were halted (see \S 5.1). Thus small planets are easier  to be scattered out of the system or hit the host star,
leaving  fewer surviving planets.

 One of the major effects that
N-body simulations can describe is the eccentricity excitation due
to mutual planetary perturbations. Such excitation may lead to some
embryos being scattered out of the system, while others may be
merged.  Thus the final number of the planets should be greatly
reduced from that of the initial embryos.
    Correlations between the number of survival planets and the averaged
mass as well as eccentricity of the planets in a planetary system
are shown in Fig.6.
 They are fitted by
 \begin{eqnarray}
 e_{\rm mean}&=&0.65\times 0.67^{N_{\rm left}},\nonumber \\
 M_{\rm mean}&=&1.28 M_J \times 0.85^{N_{\rm left}}.
 \label{nleft}
\end{eqnarray} These correlations show that, in multi-planet systems like
solar system, planets basically have relative lower averaged masses
and eccentricities than those with single or few planets.
Qualitatively, this law is easy to be understood. In order to
achieve a longer orbital crossing time in a  multi-planet system,
either the  eccentricities of the planets must be small, or the
planets should have large mutual separation scaled by their Hill
radii, thus smaller planetary  masses is helpful to  achieve a
stable system(Zhou et al. 2007).

\section{Comparing with Observations}

 To compare our results with the observations, we set the mass of
the disk ($f_g$) and its depletion timescale ($\tau_{\rm disk}$) as
two free parameters. We take  11 values of $\tau_{\rm disk}$ evenly
ranged from 0.5 Myr to 5Myr in a logarithm scale(Haisch et al.
2001).  For each $\tau_{\rm disk}$, we choose  $f_g = 0.3,1,3,10$,
and execute  6,10,5,3  simulations, respectively,   to fit for the
Gaussian distribution of
 $\log{(M_{disk}/M_{\odot})}$ with a mean value  of $\mu=-1.66$ and a standard
deviation $\sigma=0.74$,  according to  the observations of
Taurus-Auriga (Mordasini et al. 2009a). So totally we execute
$11\times24=264$ runs of simulations.

 We  also make some statistical
plots from the observed systems{\footnote{http://exoplanet.eu}}. To
show planets that may be not observable yet, we distinguish planets
of our simulations with detectables (with the induced stellar radial
velocity $V_r \ge 3~$m s$^{-1}$) and undetectables  ($V_r < 3~$m
s$^{-1}$). Taking the mean value of $\sin i=0.6$, among the 1437
survival planets,  959 planets ($66.7\%$  of total ) from our
simulations are undetectable, which contains a large number of small
planets ($<1M_{\oplus}$).

\subsection{Semimajor Axis Distributions}

Fig.7a and b show the  semimajor axis distribution from both the
observations and the simulations. The observed  distribution shows
two peaks(Fig.7a).
 (1) At 1-3AU. This is roughly the
snow line (where water is frozen) of the system, where planet cores
may be stalled under type I migration (Kretke \& Lin 2007, Ida \&
Lin 2008), subsequent accretion of gas makes them giant planets. (2)
At 0.04-0.06AU (or 3-5 days). This is roughly the inner edge of the
gas disk, where planets under type II migration will be stopped (Lin
et al. 1996). From our simulations there are lots of planets beyond
3AU, thus the lack of planets at $>3$ AU is due to observational
bias.

Besides, we show that there is an extra pile-up of planets: (3) at
around 0.2AU (or around 30 days), which has  not been  revealed by
observation yet. This location corresponds to the inner edge of the
MRI dead zone ($a_{\rm crit}$) where
 small planets under type I migration may be halted.
However,  as the location of  $a_{\rm crit}$ (See Eq.[\ref{aice}])
moves inward, in the case of short $\tau_{\rm disk}$, its migration
speed is faster than that of the  type I migration,  therefore it is
difficult to halt these planets near $a_{\rm crit}$. When $\tau_{\rm
disk}\sim~ 1.5 $Myr, $\frac{da}{dt}\mid_{\rm mig,I}\sim\frac{da_{\rm
crit}}{dt}$, so if $\tau_{\rm disk}>1.5$ Myr, the mechanism  of
halting small planets near $a_{\rm crit}$ works. After $\tau_{\rm
disk}$, the migration rate is reduced  due to gas depletion. In fact,
 $a_{\rm crit}\sim 0.2$ AU at $t=2\tau_{\rm disk}$. This
explains  the accumulation of planets near $0.2$AU.  Some small
planets were scattered into $0.04$AU.

Our simulations also show  that,  the majority  ($70\%$) of giant
planets is located in $1-10AU$. Most of them are  massive GPs and
experienced type II migration. Due to the depletion of the gas disk,
they can not migrate to the proximity of the star, hence they halted
outside $1$ AU.   Due to observational bias of radial velocity
measurements, only those massive giant planets at $< 4-5$AU  are
revealed.

\subsection{Mass Distributions}

Our simulations show mainly there are two peaks for planet masses:
(Fig.7c).

(1) At 1-2$M_J$. They are Jupiter-sized giant planets that have
grown up with efficient gas accretion. Although  our simulations
reproduced this peak, there are not enough amount of massive planets
at $M_p \sim  5-8M_J$.

(2) At $0.1-3M_\oplus$. There are a large number of terrestrial
planets,  which have grown up mainly by mutual collisions under the
perturbations of outside giant planets. This peak has not been
revealed by the observations yet, and can be checked by future
observations of terrestrial planets.

One more small peaks are revealed by our simulations:


 (3)  Around $10M_J$.    Massive embryos ($M\sim  80M_\oplus$) had
been formed after 1Myr in disks with $f_g=10$ and only experienced
time-limited type II migration. More massive giant planets($>10M_J$)
may be born with some other mechanisms, e.g., the gravitational
instability scenario.

The desert at $10-20M_\oplus$ are Neptune-sized planets. According
to our model in Section 2.3,  the Kelvin-Helmholtz timescale are
quite short($\sim10^4$yr) for this mass regime and the runaway gas
accretion makes them become giants quickly. Only a few planets in
the inner region with $M_{\rm iso}\sim10-20M_\oplus$ survive in this
desert. Also merging from lower planet embryos is possible to form
Neptune-sized planets.

\subsection{Eccentricity Distributions}

The distribution of eccentricities  from simulations is similar to
that from observations. In our simulations, the eccentricities of
survival planets vary from $0$ to $0.84$,  while the observations
show a maximum eccentricity of about $0.9$ (Wright et al. 2009). We
fit both distributions (simulations and observations) with an
exponential decay function in the form of $P(e)de=N(e)/N_{\rm
tot}\sim \exp(-Ae)de$. Since  $\int^{1}_{0}P(e)de=1$, we have \beq
P(e) =\frac{A \exp(-A e)}{1-\exp(-A)}. \label{pe} \eeq where $A=4.2$
for observations and  $A=7.8$ for  simulations of
 planets with $V_r<3m/s$, as shown in  Fig.7e and Fig.7f.
The larger value of $A$ from simulations indicates a  steeper
slope, showing more planets with small eccentricities are still not
detected, as shown in Fig.8d later.

\subsection{Correlation Graphes between $a,e$ and $M_p$}

Figure 8 presents the  correlation  diagrams of $a,e$ and $M_p$ of
our simulations (right),
 with a comparison to the observational data (left).
 Our simulations reproduced all the three correlation plots
quite well.  Especially there is a gap in the $M_p-e$ plot,
indicating a planet desert  ($0.005\sim0.08 M_J$) depends on the
 eccentricity (Fig.8c \& d).
 Fig.8d also shows a tendency that giant planets ($M_p>10M_\oplus$)
 tend to have lager eccentricity on average. To show this more clearly,
 we reproduce the eccentricity-semimajor axis correlation plots and the eccentricity distribution
 plots
 for  giant planets ($M_p>10M_\oplus$) and terrestrial planets ($M_p < 10M_\oplus$) in Fig.9.
 Giant planets have average eccentricities  $\sim 0.15$ at all locations
  (except $>10$AU), while terrestrial planets have  average eccentricities  $\sim 0.05$.
  Although both of the e-distribution can be fitted by exponential law in Eq.(\ref{pe}),
  their coefficients A  are different : 7.78  for $M_p>10M_\oplus$ and
  21.0  for $M_p < 10M_\oplus$, indicating small mass planets tend to have smaller
  eccentricities.

 To understand the $M_p-e$ desert, i.e., very few planets with
masses $0.005\sim0.08 M_J$ have eccentricities larger than
$0.3-0.4$, we plot the $e$-damping timescales $\tau_{edap}$ as a
function of planet masses at 0.3, 1, 3 and 10AU in Fig.10. For a
planet, when $M_p < M_{I,II}$, $\tau_{edap}$ is calculated by Eq.
(\ref{edpI}) using a mean eccentricity $e=0.05$ and $\Sigma_0=280$ g
cm$^{-2}$. Otherwise $\tau_{edap}$ is calculated by Eq.
(\ref{edpII}) with viscosity $\alpha_{\rm dead}=10^{-4}$. Due to Eq.
(\ref{edpI}), small planets have longer $\tau_{edap}$. The jumps of
$\tau_{edap}$ occur at $M_p=M_{I,II}$. $\tau_{edap}$ keeps
horizontal until the mass of the planets becomes comparable to the
disk mass ($M_p> 2\Sigma_p a_p^2 =21 M_\oplus (a/1\rm AU$)). In the
braking phase of type II migration, more massive planet have a
longer $e$-damping timescale according to Eq. (\ref{edpII}).
Assuming an effective damping of eccentricity if $\tau_{edap}$ is
less than $\sim$ 1Myr, we obtain a mass regime $0.005\sim0.08M_J$
which is consist with the $M_p-e$ desert from both observations and
our simulations in Fig.8c \& d. As shown in Fig.10, $\tau_{edap}$ of
massive GPs is in the horizontal region or beyond this,  so
generally they have longer $e$-damping timescales than those of the
TPs. Therefore, the mean eccentricity of TPs is smaller than that of
GPs.


\section{Simulations in Three-Dimensional Model}

 In all the above simulations, we adopted coplanar planetary model (2D).   When the orbital inclinations of the planets  (3D) are
included in the simulations,
 their final orbital characteristics of the planet systems  may be changed due to the
different collision timescales between 2D and 3D simulations (Chamber, 2001).
In this section, we reveal the effects  of including orbital inclinations  with some more simulations.
We executed 20 runs in  3D model, setting the same initial conditions with those in 2D model
except that  the initial inclinations are set as 1 degree for all
embryos, with randomly chosen. The disk parameters are also set as a standard value in our simulations:
$f_g=1, \tau_{\rm disk}=2$Myrs and $\alpha_{\rm dead}=10^{-4}$.

\subsection{Collisions with Different Masses}
During the whole evolution period we simulated (10Myrs),   there are    606 collisions in the 20 runs of 2D simulations,
while 471 collisions occurred in the corresponding runs  of 3D model.
In  the 2D runs, the number of embryos colliding with the host star or
being ejected out of the system is  120, comparing with the number of 231 in 3D
runs. It is consistent with the result of Chambers (2001).   
As seen in Fig.11a,  the cumulative distribution functions (CDF) of collisions in two models
show that most collisions in 3D runs occurred  in relatively later stages. Thus
earlier collisions in 2D model produce embryos with larger masses.
To show the effects of earlier collisions to the final planetary architectures,  we
divided the collisions as three regions, as shown in Fig.11b: Region
I ($M_0<M_{\rm crit},M_1<M_{\rm crit}$), II ($M_0<M_{\rm
crit},M_1>M_{\rm crit}$) and III ($M_0>M_{\rm crit},M_1>M_{\rm
crit}$),  where
 $M_0$ represents the larger mass of the embryo before a two-body
collision,  while $M_1$ represents that after the collision,
 $M_{\rm crit} =4M_\oplus $ is the critical mass
beyond which the efficient gas accretion sets in (see discussions below Eq.[\ref{mcrit}] ).

{\em Region I}: Most collisions (2D: $\sim75\%$, 3D: $\sim 85\%$)
occurred  in this region. These collisions only influence the
Earth-like planets(TPs, see \S 4.2). Due to the shorter collision
timescale in 2D model, embryos have larger masses on average during
early time and experience faster type I migrations and $e$-damping
according to  Eq[\ref{migI}] and Eq[\ref{edpI}]. Therefore the TPs have a
smaller mean eccentricity (c.f., $e_{\rm mean}=0.034$ for 2D and $e_{\rm mean}=0.047$
for 3D) and semimajor axis (see Fig.11c).  As a results of more collisions occurred
in 2D simulations, the final systems are expected to have less number and larger masses  of TPs
than those from 3D simulation.
In fact,  we find  11 TPs left among the total 74 survived planets in the 20 runs of 2D simulations,
while it is 24 out of the total 98 survivals in the corresponding 3D simulations,
with a small  average mass (Fig.11c).

{\em Region II}: About $8.5\%$ (2D) and $8.4\%$ (3D) collisions
occurred  in this region. These collisions makes  the
embryos with masses $<M_{\rm crit}$ grow beyond the critical mass.
However,  this effect is limited by the small fraction of collisions and the slow
gas accretion rate(Eq[\ref{dmp}]) for embryos with masses slightly larger than
$M_{\rm crit}=4M_\oplus$.  Thus the differences between  2D and 3D models due to collisions in this region
can be ignored.

{\em Region III}: Roughly $\sim 16.5\%$ (2D) and $\sim 6.6\%$ (3D) collisions
occurred  in this region. These embryos can accrete gas before
collisions. As  type I migrations speed is much faster than that of type II,
the final locations of the planets are mainly determined by their planets cores.
As planet cores  in 2D model undergo more collisions, they have bigger masses thus a fast migration speed, so their final average semimajpr axis is small than that in 3D (Fig.11d).  As inner region has less gas to accretion, $M_{\rm gas} \propto \Sigma_g a R_H$, while $R_H \propto (M_p/3M_*)^{1/3} a$ and $\Sigma \propto a^{-1}$), the final giant planets in 2D simulations have less masses(Fig.11d).
 As the type I eccentricity damping rate is much small than that of type II (Fig.10),  and from Eq.(\ref{edpI}), $\tau_{\rm edap,I}\sim a^{1/2}M_p^{-1}$, thus larger masses of planets in 3D simulations have  shorter $\tau_{\rm edap,I}$, thus
 the mean eccentricity ($e_{\rm mean}\approx0.086$ by simulations) in 3D model is smaller than that ($e_{\rm mean}\approx0.15$) in  the 2D model.

\subsection{Differences in Statistics}
As  most of our results in \S 5 are in statistics, we focus on
the statistical differences between 2D and 3D two models. The statistics
of semimajor axes and masses of planets in two models are presented
in Fig.12. Some peaks and deserts in 2D model are reproduced
in 3D model, such as the peaks at $a_{\rm crit} \sim2$AU, $M_{\rm
crit}$ and $1-2M_J$. However,  There are still some different characteristics, especially
the planet deserts at around $0.1$AU and $10M_\oplus<M_p<0.1M_J$ in 3D
model.  The lack of planets around $0.1$ AU in 3 D model may be due to the prolonged  collision timescale, a much longer time of integration may results similar qualitative results  with 2D model.
The  deficit of planets with masses  $10M_\oplus<M_p<0.1M_J$ in 3D
model make the planet desert in \S 5.2  more obvious.

Fig.13 shows the orbital inclination distribution  for the survival
planets in the 20 runs of 3D simulations. As one can see,  most of
the planets remain in $<2$ degrees, only few planets with smaller
masses have inclinations $\sim 10$ degrees.


To  summary, we point out that the differences between 2D and 3D simulations are not much, most of statistical results in 2D are
qualitatively credible.

\section{Conclusions and Discussions}

Based on the standard core accretion model of planet formation, we simulated the final
assemble of planets by integrating the full N-body equations of  motions.
The stellar mass is fixed at $1M_\odot$.
The circumstellar disk is assumed to have an
  effective viscosity parameter $\alpha_{\rm eff}$ (Eq.[\ref{alpeff}]),
 and undergoes an exponential decay in a timescale of $\tau_{\rm disk}=0.5-5$Myr.
Initially embryos with isolation masses larger than $0.1M_{\oplus}$
(Eq.[\ref{msio}]) are put in each system.
 For embryos below (or above) the critical mass defined
in Eq. (\ref{MIII}), they will undergo type I (or type II, resp.) migration.
The type I migration of embryos will be stalled near the inner edge of
the MRI dead zone defined in Eq.(\ref{acrit}). The inner edge of the disk is set as
$0.05$AU where giant planets will be stalled under  type II migration.
The equations  governing the embryo's  motions are shown in Eq. (\ref{eqf}).

 We have investigated the influences of different
parameters(viscosities, disk masses, disk depleted timescales) on
the final architectures of the planetary systems. {\em Disk
viscosity} affects the planet locations by determining their
migration speeds. In the regime of $\alpha_{\rm dead} \in
[10^{-4},10^{-2}]$ , planets in disks with small viscosities migrate
fast under type I planet-disk interactions, which results in a small
average semimajor axis.  For a moderate disk viscosity,  the
planetary system shows a moderate averaged locations (Fig.3a). The
average eccentricity does not show obvious correlation with the disk
viscosity. {\em Disk mass} affects the system through the initial
core masses. Large planet  cores can be formed in the outer region
of a massive disk, so that giant planets are easier to form.

{\em Disk depletion timescale} ($\tau_{\rm disk}$) plays an
important role in final planetary masses and orbits. For short-lived
disks with $\tau_{\rm disk} \le 1$Myr,  the forming planets have
small masses ($70-200 ~M_\oplus$), with large average semimjor axes
($5-6 $AU).  This indicates that  most of the  surviving planets are
formed  in distant orbits, while planet migrations did not affect
their orbital architectures yet.  Due to the short lifetime of the
disk,
 their average eccentricities ($\sim 0.15$) are relatively big.
 Planetary systems with a long-lived disk ($\tau_{\rm disk}>1 $Myrs) tend to have large
planetary masses ($\sim 1$ Jupiter mass ), with low averaged
eccentricities ($\sim 0.1$) due to the long time disk-damping. The
average semimajor axes of these planets  are ranging from   4 AU to
2 AU, indicating inward migration  indeed plays an  important role
in making planets in close-in  orbits.

 Comparing the simulations to others, we find
 the number of planets  being trapped in mean motion resonances (MMRs) are smaller than
 those obtained, such as in Terquem \& Papaloizou (2007).
  Instead, we get a large amount of
 planet pairs that have the history of passing periastron alignment (See Table 2, with the difference
 of their longitude periastron librating at $0^o$).
  The reason that we did not observe many survival planets in MMRs may be due to the fact that,
 most of our survival planets are giant planets.
 The choice of a small disk viscosity ($\alpha_{\rm dead}=10^{-4}$) makes the type II migration  too slow
   to make a significant convergent MMR trapping.
 For terrestrial planets,  the speed of type I migration is faster than that of type II migration of
 the giant planets in outside orbits, so they are not easy to be  trapped into the MMRs of giant planets as well.

Statistics of 264 simulated systems reproduced
 qualitatively the main features on the planet masses and orbital
 parameters for the observed exoplanetary systems.
 If we classify the planets into two major categories:
 giant planets (GPs), including massive GPs ($M_p \ge 30M_\oplus$)
 and Neptune-sized GPs ($10M_\oplus  \le M_p<30M_\oplus$);
 terrestrial planets (TPs), including Super-Earth TPs ($1M_\oplus  \le M_p<10M_\oplus$)
 and Sub-Earth TPs($M_p<1M_\oplus$), then
 results from our simulations   have the following {\em implications} on these planets.

{\em Occurrence rates of planets}.
  The ratio of GPs relative to TPs is low, i.e., 514 to 923 (or 36\% to 64\%)
  in our total 1437 survival planets (table II). This
   is mainly due to the fact that only those massive embryos can accrete
   sufficient gas to form GPs. Planets with smaller masses are easier to be scattered
   out by N-body interactions, which is the  major difference with the population synthesis simulations of single-planet systems.
 Moreover,  there exist some correlations between the
survival number of planets  and the average eccentricity (or average
planet mass) of a planetary system, i.e., a planetary system with
more planets  tends  to have smaller planet  masses and orbital
eccentricities(Fig.6 and Eq. [\ref{nleft}]). These correlations are
consistent with the  stability of the system, i.e., a system with
planets in less eccentric  orbits and with larger mutual separation
scalled by their Hill radii  tends  to have longer crossing
timescale, thus is more stable (Yoshinaga et al. 1999, Zhou et al.
2007).

 {\em Locations of giant planets }(GPs). Most  (298, $\sim 58\%$ of total 514 ) of   GPs  locate
  at orbits with semimajor axes
 1-10 AU , with only $\sim 35 \% $ (182)  at inner orbits $<1$ AU.
 This may be linked  with   the snow line ($\sim 2-3$AU) of the system.
 Due to the surface density  enhancement
of about $3-4$ (Eq. [\ref{msio}]), isolated masses beyond the snow
line are larger thus they are easier to accrete gas and become giant
planets.
 Neptune-sized GPs are formed in our simulations also, with
 its amount being much smaller than  massive GPs ( $\sim7.5 \% $, 42 of 1437),  and they
  mainly locate at two parts of the systems:  either at $\sim 10$AU, which were
 scattered out by first formed giant planets (like the formation of
 Uranus and Neptune),  or in the inner
 edge of the gas disk ($0.05$AU), which seems to be stalled there under type II migration.

  {\em Locations of terrestrial planets }(TPs).
  TPs are almost evenly distributed, except a group lying at  1-2
  AU(Fig.4d, for $f_g=1$), just inside  the snow line($\sim 3$AU).
 Also  there is a  slight pileup of planets (both GPs and TPs) at   $0.2 -0.3$AU (30-50days),  which may correspond to the inner boundary of the MRI dead zone.
  According to Kretke \& Lin (2007) and Morbidelli et al. (2008),
super-Earths are easier to be stalled there from type I migration,
and grow up to giant planets by gas accretion.

 {\em Eccentricities of planets}.  There are very few planets with masses
$0.005\sim0.08 M_J$ that have eccentricities larger than $0.3-0.4$,
 The average eccentricities
 of  giant planets are larger ($\sim 0.15$) than those of the
  terrestrial planets($\sim 0.05$, Fig.9).
According to our  simulations, the underlying mechanism is the
relatively long $e$-damping timescale of massive planets due to the
gas disk, especially when the planet mass is larger than the inner
disk mass, see Eq.(\ref{migII}) and Eq.(\ref{edpII}).

 We compared our results with some 3D simulations in \S 6.  The qualitative differences between 2D and 3D models are not big,
indicating our conclusions based mainly on 2D simulations are still reliable.

 Now let's compare above conclusions  to the five new observational signatures  that
 stated in \S 1.   According  to our
 simulations,  260 runs in total 264 ones result in multi-planet systems, which rates to  $98\%$,
 comparing with that from the observations ($>28\%$). Planets in multiple systems have smaller eccentricities
 than single planets, which is  revealed clearly in Fig.6.
 For signature-2, we did not classify our $a,e$ and $M_p$ distributions with single and multiple systems, as this
 classification has some uncertainty concerning  undetected planets. While Fig.9a  supports
   signature-3, massive planets  seem to have larger eccentricities than  those of
   smaller mass planets. The implication of this is interesting. As most of the
   observed exoplanets have Jupiter-sized planets in elliptical orbits, we can expect more
   planets with small masses in near circular orbits  of multi-planet system,
    like terrestrial planets in  the solar system.
   Signature-4, i.e., massive planets tend to
   have larger eccentricities,  is revealed by our simulations (Fig.9b).
   For signature-5, as we fix the stellar mass
   at $1~ M_\odot$,  it  can not be tested in the present work. We will investigate this
   correlation in future works.

However, due to some {\em limits and uncertainties} of the
parameters we chose in this model,  the conclusions obtained in the
paper may be limited. For example,   we investigated  planet
formation around  1 $M_\odot$ stars only, although with different
sizes of the solid and gas disks. Not only the disk mass but also
the accretion rate as well as metallicity have correlation with the
mass of the host star.   For more broader parameters, the occurrence
ratio of planets between TPs and GPs
    may need further investigations.

    One of the major uncertainties arises from the initial
 masses and distributions of the embryos, which are the building blocks of planets.
 In this paper, we adopt the assumption that most of the embryos have already
 cleared their nearby heavy elements and achieved their isolation masses. This assumption
  is based on the full N-body simulation of planet growth (Kokubo \& Ida 1998), and
 will be too ideal when type I migration
 of embryos are taking into considerations.
 Also, initially there might  be some smaller
embryos between giant planets so that they may be scattered
 through planet scattering and become planets like
Uranus and Neptune.

 The second major uncertainty  comes from our poor knowledge of  type I migration of
 the embryos. The speed  (and even the direction) of type I migration will
affect the number and locations of survival terrestrial planets,
especially planets in habitable zones (e.g.,  Ogihara  \& Ida 2009,
Wang \& Zhou 2010). We will do more investigations to that end in
our forthcoming works.

Thirdly, our simulations mainly focus on the stage that the gas disk
is present and will be depleted exponentially. Further evolution due
to the effect of a planetesimal disk was not included. During the
later stage of planet formation when the gas disk is totally
depleted, planet-planetesimal interactions may damp the
eccentricities of planets through dynamical frictions, and may
induce migrations through angular momentum exchanges with embryos
and scattered planetesimals (Fernandez \& Ip 1984; Malhotra 1993).
In the solar system, numerical simulations show that, Jupiter will
drift inward, while Saturn, Uranus and Neptune may migrate outward,
resulting in a divergent migration. During the migration, the cross
of 2:1 MMR between Jupiter and Saturn
  excites the eccentricities of four giant planets (Tsiganis et al. 2005).
Such evolution may occur on the timescale of at least hundreds of million years, which
is beyond the ability of our simulations.

Furthermore, we ignored the tidal effect between a host star and
close-in planets. The tidal dissipation of the star-planet system
may damp the eccentricity of the planets in  close-in orbits in a
Gyr timescale. Also, the gravitational potential of the gas disk was
not included in our model. During disk depletion, the sweeping of
secular resonance through inner region may  excite the
eccentricities of inner orbits, thus  induce further mergings of
terrestrial planets (Nagasawa et al. 2003).

Although with  many restrictions, the present paper, aiming at the
statistics of final assembling of planetary systems in the standard
formalism (which includes  type I and  type II migration, gas
accretion, etc.), reproduces most of the observed orbit signatures.
 Thus we think that the  predictions by the simulations are
 helpful for guiding future planet detections.

{\bf Acknowledgements}
  Zhou is very grateful for the hospitality of Issac Newton institute
during the program of 'Dynamics of Discs and Planets'.
 This work is supported
by NSFC (10925313,10833001, 10778603), National Basic Research Program of
China (2007CB814800), and a Fund from the Ministry of Education,  China (20090091110002).

\clearpage

\begin{table}
\begin{center}
\caption{Notations used in the paper, $\tilde{M}_*=M_*/M_\odot,
\tilde{r}=r/1{\rm AU}$,
\label{tbl-n}}.
\begin{tabular}{ll}
\tableline \tableline
$T=280 \tilde{M}_* \tilde{r}^{-1/2}$ & disk temperature by stellar radiation \\
$c_s=1.2 {\rm km ~s}^{-1} \tilde{M}_*^{1/4} \tilde{r}^{-1/4}$ & sound speed in the ideal gas with adiabatic index $\gamma=1.4$  \\
$v_k=29.8 {\rm km ~s}^{-1} \tilde{M}_*^{1/2} \tilde{r}^{-1/2}$ & orbital speed of Kepler motion \\
$h=r c_s/v_k =0.047 \tilde{r}^{1/4} r  $ & veritical scale height of the gas disk \\
$\nu=\alpha c_s h$    &  disk viscosity by $\alpha$ model \\
$\beta= -\ln \Sigma_g /\ln a$ & slope of the gas disk surface density \\
\tableline
\end{tabular}
\end{center}
\end{table}

\begin{table}
\begin{center}
\caption{Statistics of the survival planets in 264 runs of
simulation.\label{tbl-1}}.
\begin{tabular}{lrrrrr}
\tableline\tableline
Planet  Mass &  No.   &  Passing PAs &  Passing MMRs  & Final PAs  & Final MMRs \\
\tableline
$M_p>30M_\oplus$ & 473 & 193 & 2 & 78 & 0 \\
$10M_\oplus<M_P<30M_\oplus$ & 41 &18& 5 & 12 & 0 \\
$1M_\oplus<M_P<10M_\oplus$ &  344 & 297 &  2 & 148 & 0 \\
$M_p<1M_\oplus$ & 579 & 306  & 0 & 82 & 0\\
Total & 1437 & 814 & 9 & 320 & 0 \\
\tableline
\end{tabular}
\end{center}
Notes: PA means periastron alignment, MMR means mean motion resonance. Passing means at some stage of evolution (including the final time),  planets are trapped either in PAs or MMRs.
\end{table}

\newpage

\begin{figure}
 \vspace{0cm}\hspace{0cm}
 \epsscale{1}  \plotone{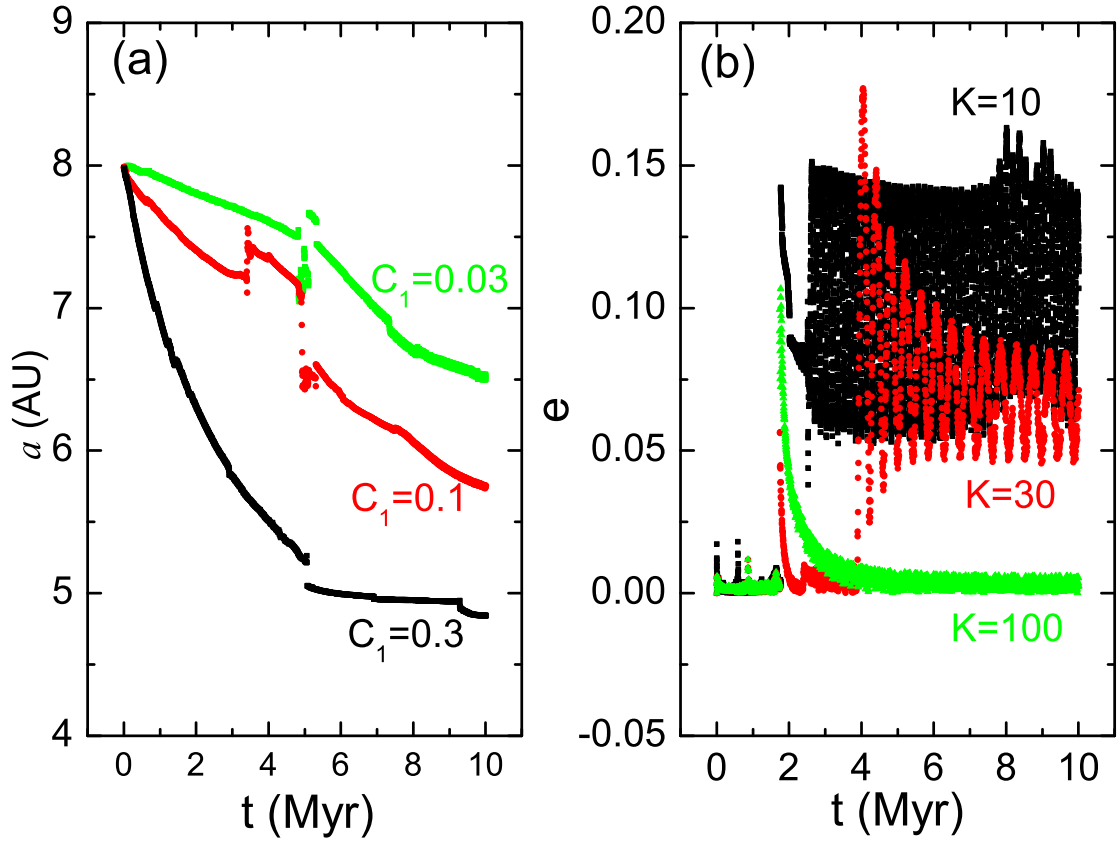}
 \vspace{0cm} \caption{Some tests are run to determine the appropriate values of type I migration reduction factor
 $C_1$ in Eq.(\ref{migI}) and the eccentricity-damping rate $K$ in Eq.(\ref{edpII}).   (a) Migrations of
 a planet under different $C_1$ .   (b)  Eccentricity evolutions of
 a planet under different $K$ .
   \label{fg1}}
\end{figure}

\begin{figure}
 \vspace{-1cm}\hspace{0cm}
 \epsscale{1}  \plotone{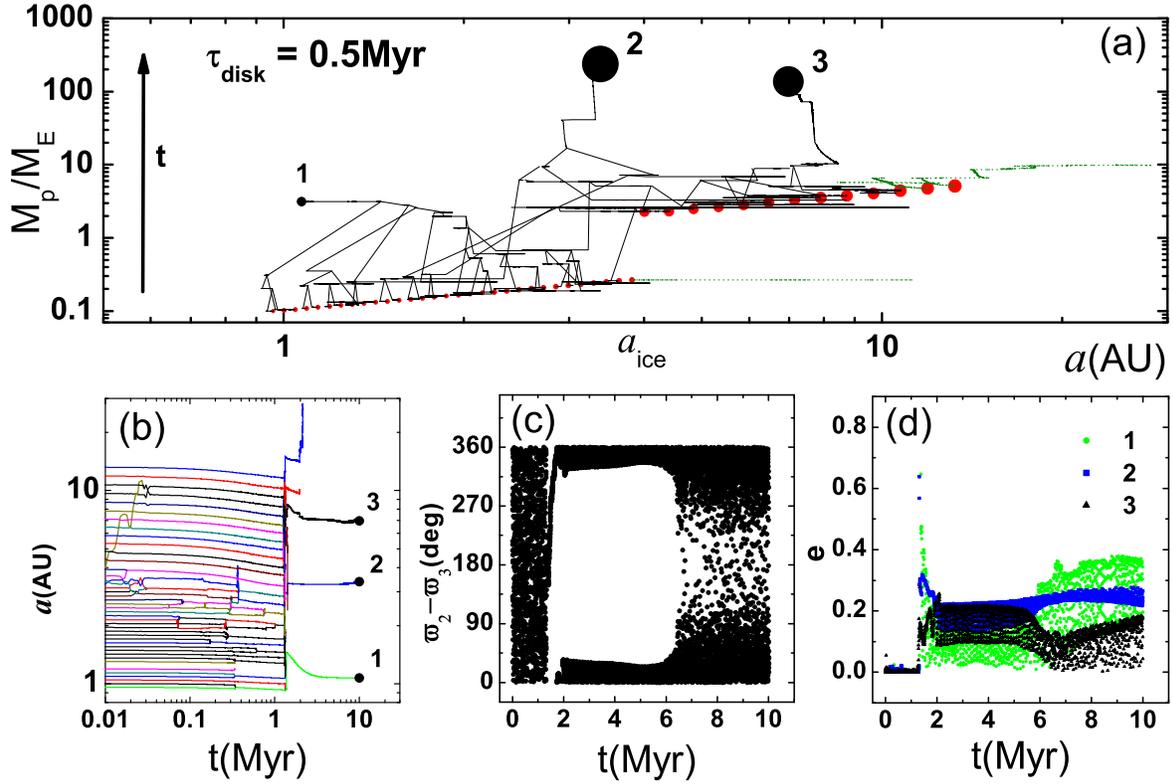}
\caption{Evolution of orbits in a typical run
 that two of the three survival planets passed  through a
 periastron alignment.  Initially 44 embryos are put with a disk depletion timescale $\tau_{disk}=0.5$ Myr. Three planets are survival with masses
 $3.1 M_\oplus$, $238.2 M_\oplus$, $137.6 M_\oplus$,
 and semimajor axes  1.07AU, 3.38AU, 6.98AU, respectively.
  (a)Planet growth and
 evolution in mass-semimajor axis plane. The
 red dots are the initial locations of the embryos. The green dots are those with either $a>50$AU or $e>1$ (being scattered out) during the evolution.
 (b) Evolution of semimajor axes of all embryos.
 (c) Evolution of periastron alignment angle ($\varpi_2-\varpi_3$).
 (d) Evolution of eccentricities for the three survival planets.
\label{fg2}}
\end{figure}
\clearpage

\begin{figure}
 \vspace{-1cm}\hspace{0cm}
\epsscale{1}  \plotone{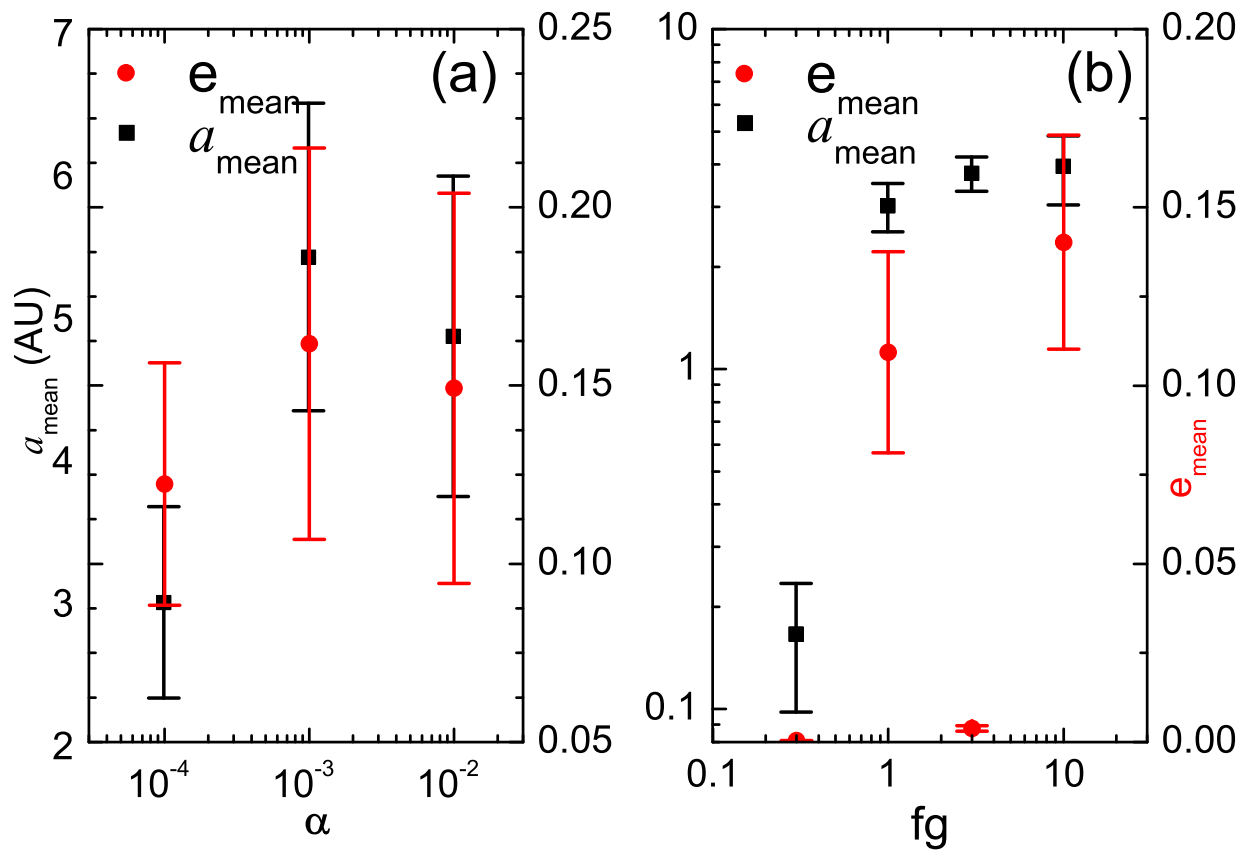} \caption{The influences of survival
system properties due to different gas disks. The left vertical axis
presents mean semimajor axis while the right one presents mean
eccentricity. (a) influence of effective viscosity parameter
$\alpha$, (b) influence of the disk mass. the error bars are the
standard deviations. \label{fg3}}
\end{figure}
\clearpage

\begin{figure}
 \vspace{-1cm}\hspace{0cm}
 \epsscale{1} \plotone{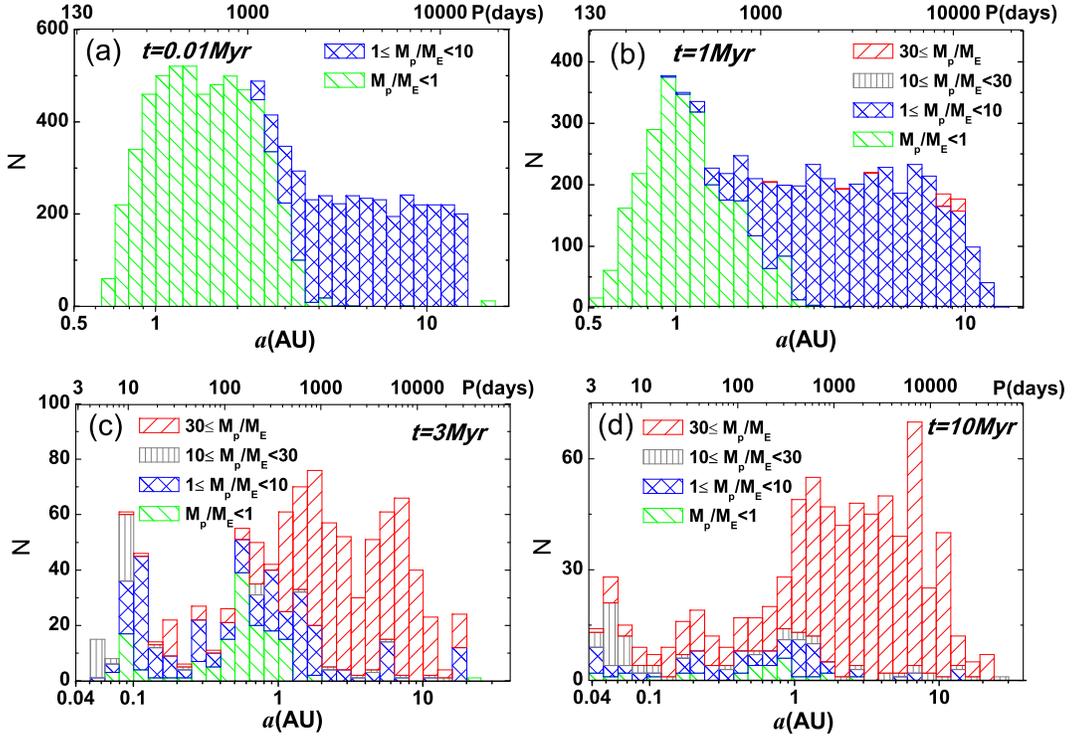}
\caption{Distributions of semimajor axes for the survival planets in
the 220 runs of simulations at different epoches:  (a) 0.01 Myr, (b)
1Myr, (c) 3 Myr, (d) 10Myr during the evolution. \label{fg4}}
\end{figure}
\clearpage

\begin{figure}
 \vspace{-1cm}\hspace{0cm}
\epsscale{1}  \plotone{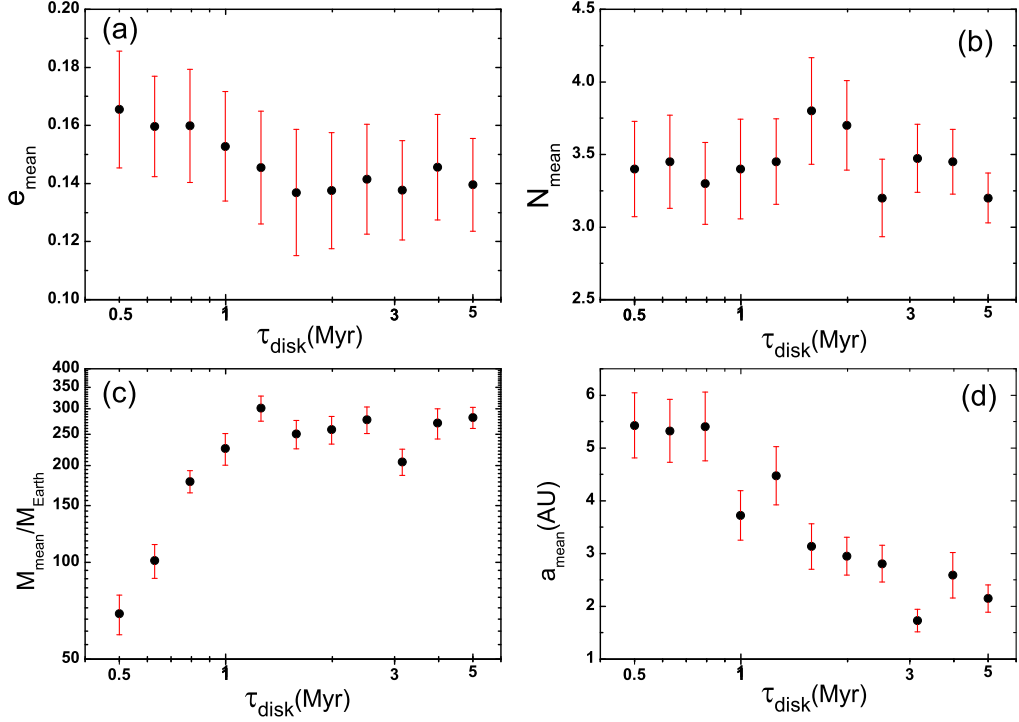} \caption{Correlations between the
parameters of survival systems and the disk depletion timescale
($\tau_{\rm disk}$) in 220 runs of simulations.  Each point is
average over 20 runs of systems with same $\tau_{\rm disk}$, with
error bars indicating the standard deviation.   Panels from  (a) to
(d) shows the average eccentricities,   numbers, masses  and
semimajor axes of the surviving planets versus $\tau_{\rm disk}$.
\label{fg5}}
\end{figure}
\clearpage

\begin{figure}
 \vspace{-1cm}\hspace{0cm}
\epsscale{1}  \plotone{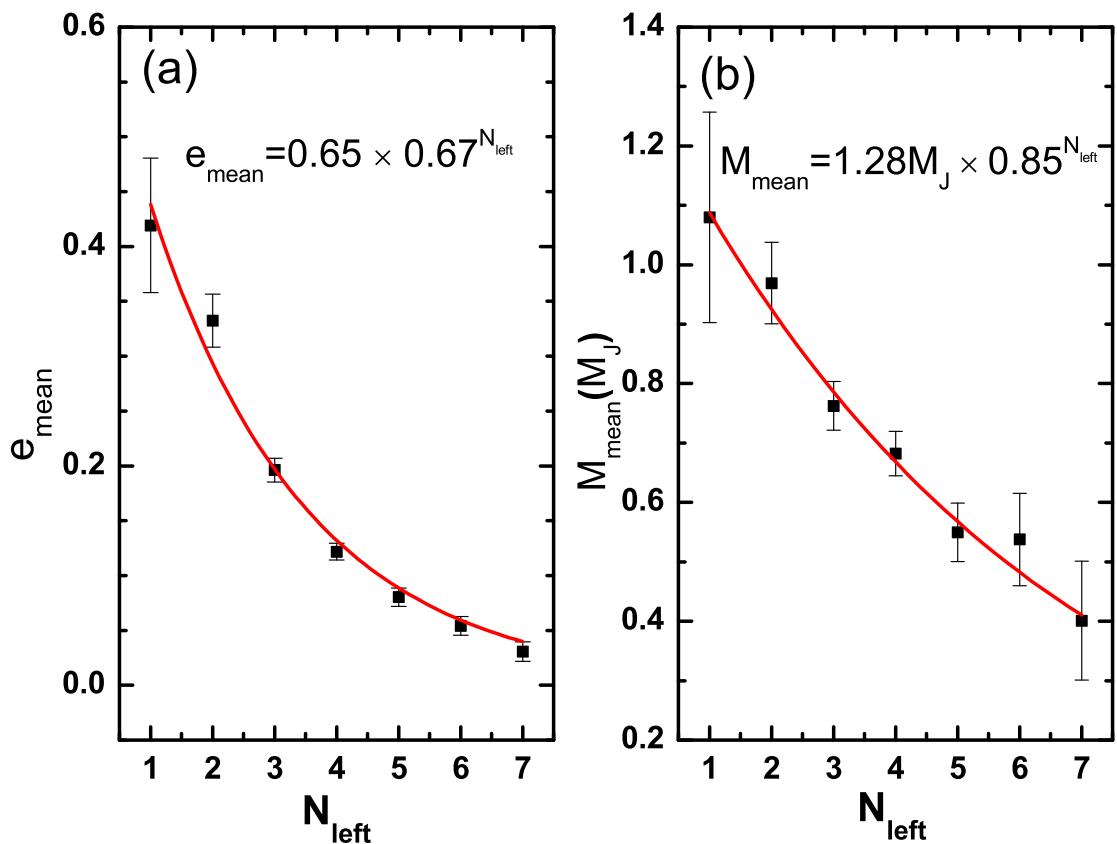} \caption{The correlations between
the survival planet number  $N_{\rm left}$ and the averaged
eccentricity $e_{\rm mean}$ (a) as well as the average planet mass
$M_{\rm mean}$ (b) in a planetary system.
 Both $e_{\rm mean}$ and $M_{\rm mean}$  are inversely
correlated with $N_{\rm left}$. The error bars are the standard
deviations. \label{fg6}}
\end{figure}
\clearpage

\begin{figure}
 \vspace{0cm}\hspace{0cm}
 \epsscale{1}  \plotone{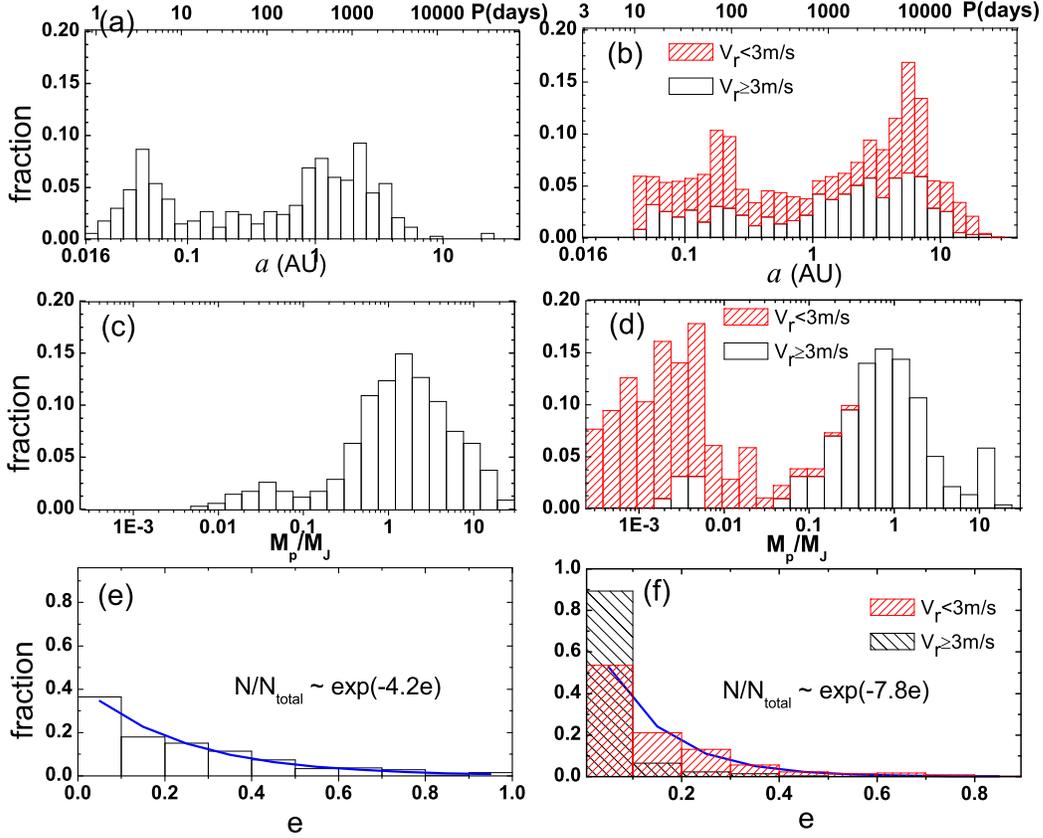}
 \vspace{0cm}
\caption{Distributions of  semimajor axes, eccentricities and planet masses ($a,e,M_p$, resp.) from observations (left column, data: http://exoplanet.eu) and  our simulations (right column). The shaded bars in  (b),(d),(f)
 show distributions of the undetectable planets (with radial velocity $V_r<3 {\rm m/s}$).
 The solid curves in (e) and (f) are fitting curves by  $N(e)/N_{\rm total}=0.1 A \exp(-Ae)/[1-\exp(-A)]$, with
$A=4.2$ for observations (e) and $A=7.8$ for simulations of planets
with $V_r>3 {\rm  m/s}$ (f). \label{fg7}}
\end{figure}

\clearpage
 \begin{figure}
 \vspace{0cm}\hspace{0cm}
\epsscale{1}  \plotone{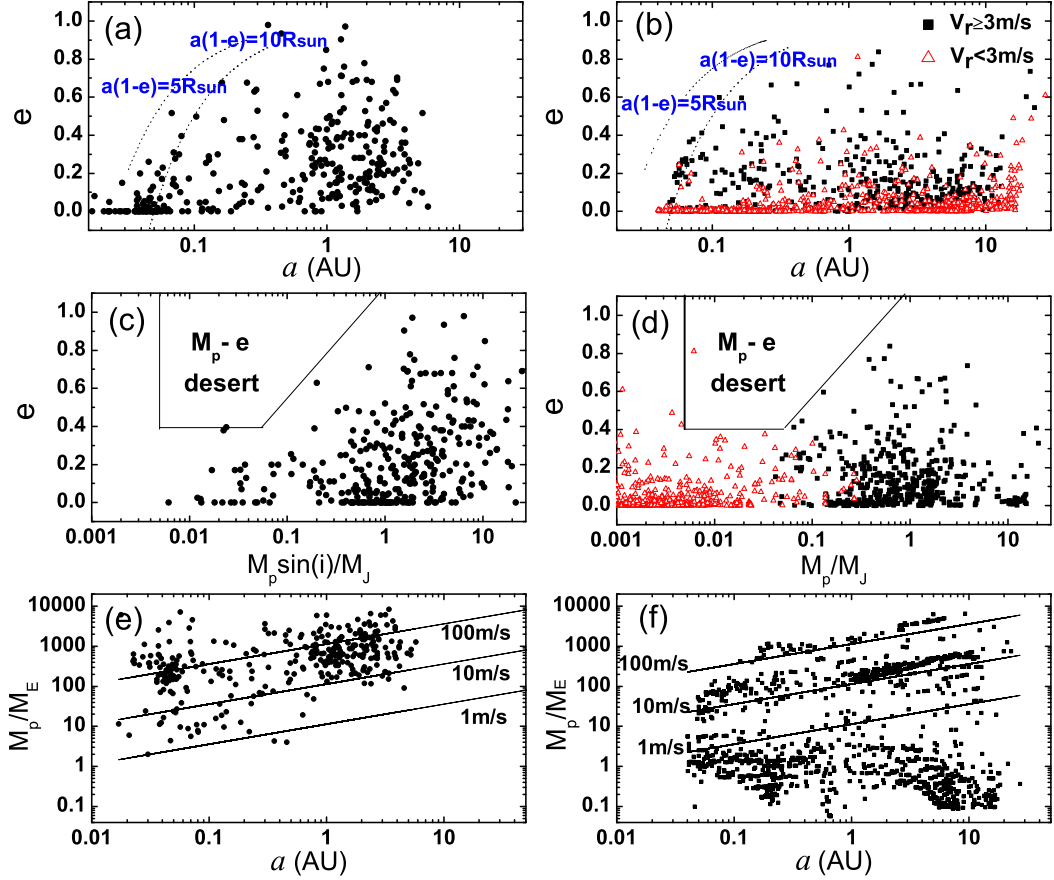} \caption{Correlation graphes from
observations(left column, data: http://exoplanet.eu)
 and our simulations (right column).
 Planets with induced stellar radial velocities $V_r < 3 {\rm m/s}$ are shown in red
 triangles.
 Solid lines in (e) and (f) show the induced stellar radial velocities.  The boxes in (c) \& (d) show the planet desert in
 the $M_p-e$ diagram.
\label{fg8}}
\end{figure}

\clearpage

\begin{figure}
 \vspace{0cm}\hspace{0cm}
 \epsscale{1}  \plotone{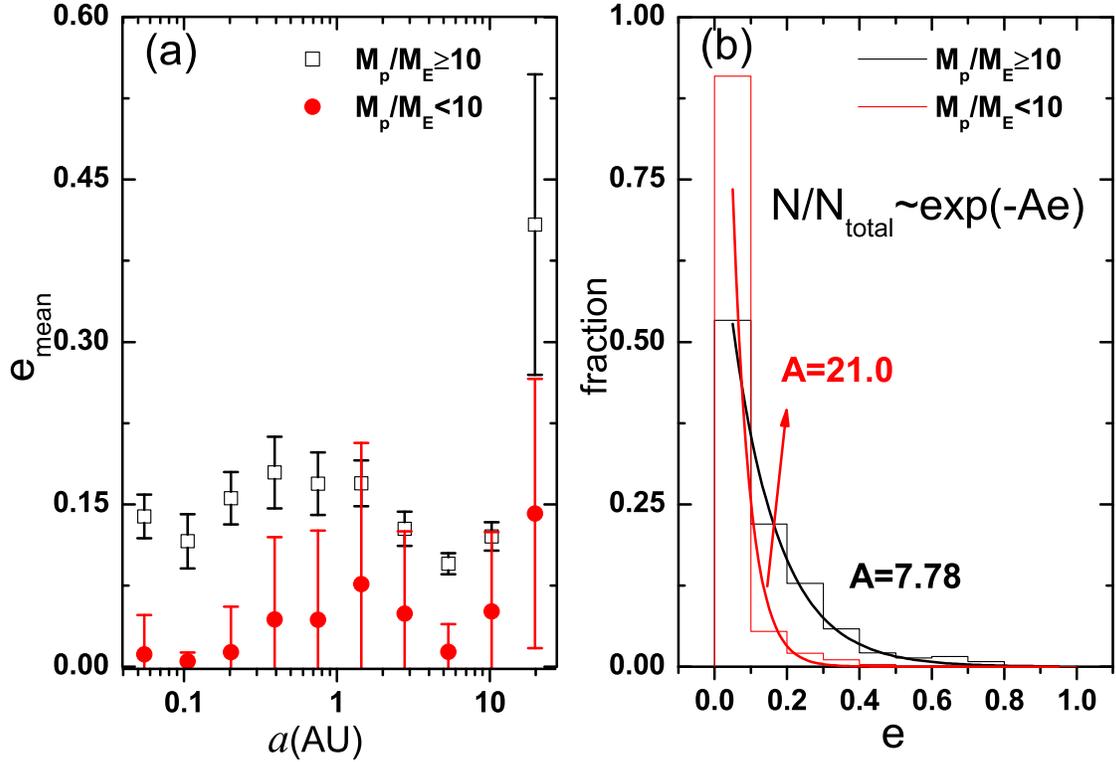}
 \vspace{0cm}
\caption{Variations of mean eccentricities
with semimajor axes (a) and distributions of
eccentricities (b) for giant planets ($M_p \ge  10M_\oplus$) and terrestrial
 planets ($M_p < 10M_\oplus$).
 Histograms in (b) are fitted
 by $N(e)/N_{\rm total}=0.1 A \exp(-Ae)/[1-\exp(-A)]$, with
 $A=7.78$ for giant planets and $A= 21.0$ for terrestrial planets.
  \label{fg9}}
\end{figure}
\clearpage

\begin{figure}
 \vspace{0cm}\hspace{0cm}
 \epsscale{1}  \plotone{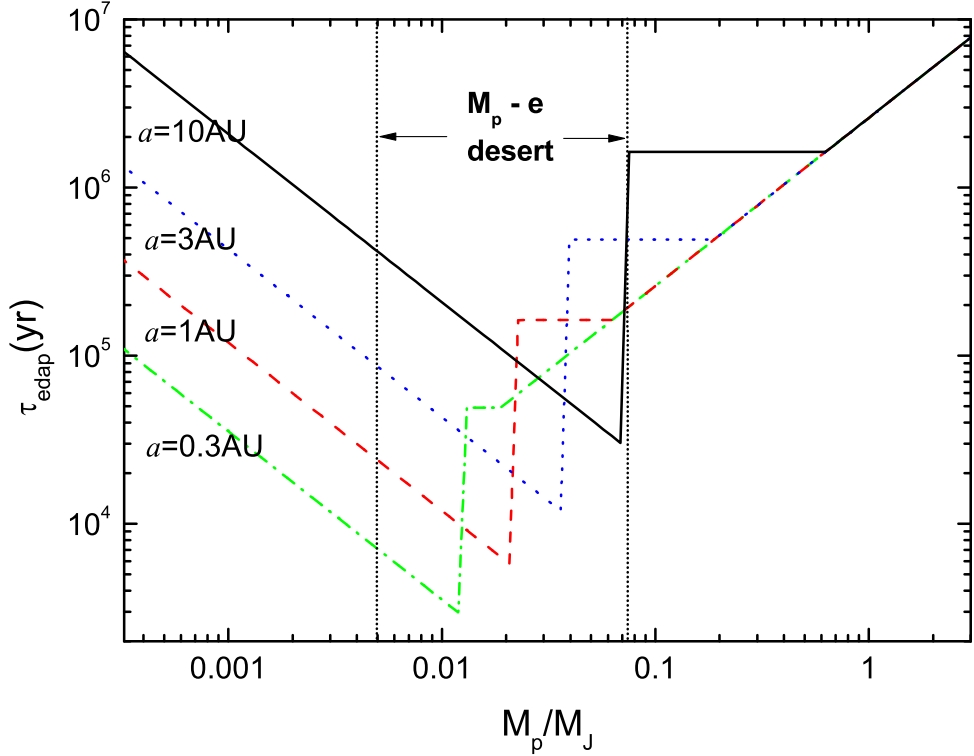}
 \vspace{0cm}
\caption{Variations of eccentricity damping timescale ($\tau_{edap}$) for different planet masses ($M_p$)
at semimajor axes 0.3AU, 1AU, 3AU and 10 AU. For those planets $M_p < M_{I,II}$, $\tau_{edap}$ is
calculated by Eq.(\ref{edpI}) using a mean eccentricity $e=0.05$.
Otherwise $\tau_{edap}$ is calculated by Eq.(\ref{edpII}). The jump
of $\tau_{edap}$ occurs at $M_p=M_{I,II}$. $\tau_{edap}$ keep
horizontal until the mass of the planet becomes comparable to the
disk mass.  The  $M_p-e$ desert in Fig. 8d  roughly corresponds to the mass regime with $\tau_{edap}$ less than $\sim 1Myr$.
 Out of this region eccentricities of planets can
hardly be damped. \label{fg10}}
\end{figure}
\clearpage

\begin{figure}
\vspace{0cm}\hspace{0cm}
\epsscale{1}  \plotone{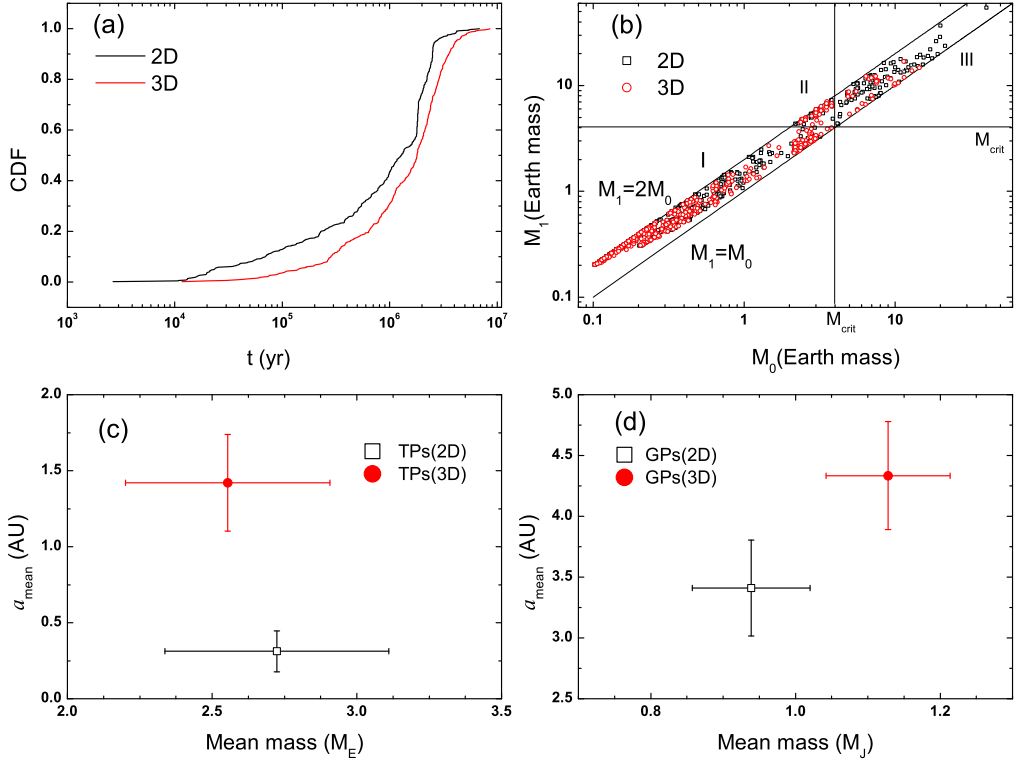}
\caption{(a) The cumulative distribution function(CDF) of collision
time. (b) The masses of embryos before collisions $M_0$ and the
masses after collisions $M_1$.(c) \& (d): Mean semimajor axis and
mass for TPs and GPs respectively. In panel (c): $M_{\rm
mean}=2.72M_\oplus, a_{\rm mean}=0.31$AU (2D) and $M_{\rm
mean}=2.55M_\oplus, a_{\rm mean}=1.42$AU (3D) for TPs. In panel (d):
$M_{\rm mean}=0.94M_J, a_{\rm mean}=3.41$AU (2D) and $M_{\rm
mean}=1.13M_J, a_{\rm mean}=4.33$AU (3D) for GPs.\label{fg11}}
\end{figure}

\clearpage

\begin{figure}
\vspace{0cm}\hspace{0cm}
\epsscale{1}  \plotone{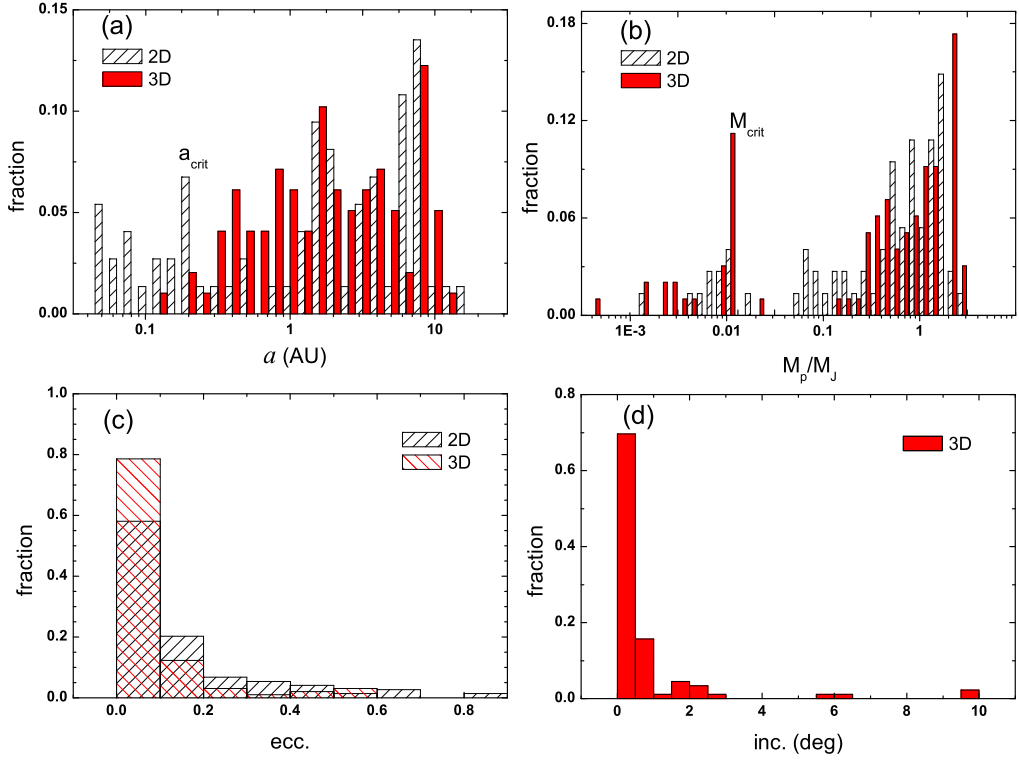}
\vspace{0cm}
\caption{ Distributions of (a) semimajor axis, (b) planetary masses,
(c) eccentricities and (d) inclinations of the survival planets in 20 runs of 2D simulations and 20 runs of 3D simulations, respectively. \label{fg12}}
\end{figure}
\clearpage

\clearpage

\begin{figure}
\vspace{0cm}\hspace{0cm}
\epsscale{1}
\plotone{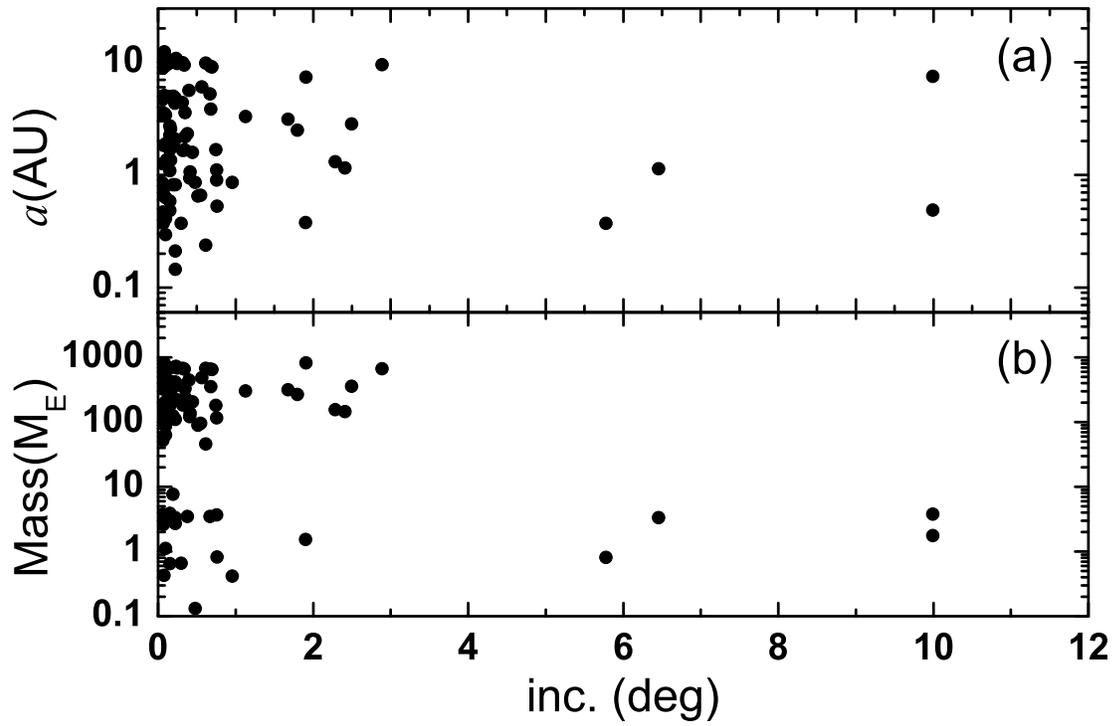}
\vspace{0cm}
\caption{Correlation graphes between inclinations with  (a) semimajor
axes, (b) masses in the 20 runs of 3D simulations, respectively. \label{fg13}}
\end{figure}


\begin{thebibliography}{99}
\bibitem[Aarseth 2003]{Aarseth}Aarseth, S. J. 2003, Gravitational N-Body Simulations, Cambridge University Press, Cambridge.
\bibitem[Alibert et al. (2005)]{Ali05}Alibert, Y., Mordasini, C., Benz,W., \& Winisdoerffer, C. 2005, A\&A, 434, 343
\bibitem[Andrews \& Williams 2005]{Andrews}Andrews, S. M., \& Williams, J. P. 2005, ApJ, 631, 1134
\bibitem[Armitage \& Rice (2005)]{Armitage}Armitage, P. J., \& Rice, W. K. M. 2005, In STScI Symp. Ser. 19, ed. M. Livio, K. Sahu, \& J. Valenti (Cambridge University Press), 66
\bibitem[Balbus \& Hawley (1991)]{BH91}Balbus, S. A., \& Hawley, J. F. 1991, ApJ,  376,  214
\bibitem[Beckwith \& Sargent 1996]{Beckwith}Beckwith, S. V. W., \& Sargent, A. I. 1996, Nature, 383, 139
\bibitem[Chambers 2001]{Chambers}Chambers, J. E. 2001, Icarus, 149, 262
\bibitem[Chatterjee et al. 2008]{Cha08}Chatterjee, S., Ford, E. B., Matsumura, S., \& Rasio, F. A. 2008, ApJ,  686,  580
\bibitem[Cresswell \& Nelson (2006)]{Cresswell}Cresswell, P., \& Nelson, R. P. 2006, A\&A, 450, 833
\bibitem[Fernandez \& Ip 1984]{FI84}Fernandez, J. A., \& Ip, W.-H. 1984, Icarus, 58, 109
\bibitem[Fischer \& Valenti 2005]{FV05}Fischer, D. A., \& Valenti, J. A. 2005. ApJ,   622, 1102
\bibitem[Gammie 1996]{Gammie}Gammie C.F. 1996, ApJ, 457, 355
\bibitem[Garaud \& Lin 2007]{GL07}Garaud, P., \& Lin, D. N. C, 2007,ApJ, 654,606
\bibitem[Goldreich \& Tremaine 1979]{GT79}Goldreich, P., \& Tremaine, S. 1979, ApJ, 233, 857
\bibitem[Goldreich \& Tremaine 1980]{GGT80}Goldreich, P., \& Tremaine, S. 1980, ApJ, 241, 425
\bibitem[Gullbring et al. 1998]{Gull98}Gullbring,E. et al. 1998, ApJ.  492, 323,
\bibitem[Haisch et al. (2001)]{Haisch}Haisch, K. E., Lada, E. A., \& Lada, C. J. 2001, ApJ, 553, L153
\bibitem[Hayashi 1981]{hay1981}Hayashi, C. 1981, Prog. Theor. Phys. Suppl., 70, 35
\bibitem[Ida \& Lin (2004a)]{IL04a}Ida, S., \& Lin, D. N. C. 2004a, ApJ, 604, 388
\bibitem[Ida \& Lin (2004b)]{IL04b}Ida, S., \& Lin, D. N. C. 2004b, ApJ, 616, 567
\bibitem[Ida \& Lin (2005)]{IL05}Ida, S., \& Lin, D. N. C. 2005, ApJ, 626, 1045
\bibitem[Ida \& Lin (2008)]{IL08}Ida, S., \& Lin, D. N. C. 2008, ApJ, 673, 487
\bibitem[Ikoma et al. 2000]{iko00}Ikoma, M., Nakazawa, K., \& Emori, H. 2000, ApJ, 537, 1013
\bibitem[Juri\'{c} \& Tremaine]{JT08} Juri\'{c}, M., \& Tremaine, S. 2008, ApJ, 686, 603
\bibitem[Johnson et al. 2007]{John07}Johnson, J. A. et al. 2007, ApJ, 670, 833
\bibitem[Kennedy \& Kenyon (2008)]{Kennedy}Kennedy, G. M., \& Kenyon, S. J. 2008, ApJ, 673, 502
\bibitem[Kley et al. 2009]{KBK09}Kley, W., Bitsch, B., \& Klahr, H. 2009, A\&A, 506, 971
\bibitem[Kokubo et al. 1998 ]{Ko98}Kokubo, E., Yoshinaga, K., \& Makino, J. 1998, MNRAS, 297, 1067
\bibitem[Kokubo \& Ida 1998]{KI98}Kokubo, E., \& Ida, S. 1998, Icarus, 131, 171
\bibitem[Kokubo \& Ida 2002]{KI02}Kokubo, E., \& Ida, S. 2002, ApJ, 581, 666
\bibitem[K\"{o}nigl 1991]{Kon91}K\"{o}nigl,A. 1991, ApJ, 370, L39
\bibitem[Kornet \& Wolf 2006]{Kornet}Kornet, K., \& Wolf, S. 2006, A\&A, 454, 989
\bibitem[Kretke \& Lin, 2007]{Kretke}Kretke, K. A., \& Lin, D. N. C. 2007, ApJ, 664, L55
\bibitem[Kretke et al. 2008]{Kretke08}Kretke, K. A., Lin, D. N. C., Garaud, P., \& Turner, N. J. 2009, ApJ, 690, 407
\bibitem[Laughlin et al. 04]{LSA04}Laughlin, G., Steinacker, A., \& Adams, F. C. 2004, ApJ, 608, 489
\bibitem[Lee \& Peale (2002)]{Lee}Lee, M. H., \& Peale, S. J. 2002, ApJ, 567, 596
\bibitem[Lin, Bodenheimer \& Richardson (1996)]{Lin}Lin, D. N. C., Bodenheimer, P., \& Richardson, D. C. 1996, Nature, 380, 606
\bibitem[Lin \& Papaloizou 1979]{Lin}Lin, D. N. C., \& Papaloizou, J. C. B. 1979, MNRAS,  188, 191
\bibitem[Lin \& Papaloizou 1985]{Lin}Lin, D. N. C., \& Papaloizou, J. C. B. 1985, in Protostars and Planets II, ed. D. C. Black \& M. S. Matthew (Tuscon: Univ. Arizona Press), 981
\bibitem[Lovis et al. 2010]{Lovis10}Lovis,C.,  S{\'e}gransan, D., Mayor, M., et al. 2010, A\&A, (submitted)
\bibitem[Malhotra 1993]{Mahaltra}Malhotra, R. 1993, Nature, 365, 819
\bibitem[Mizuno 1980]{miz80} Mizuno, H.  1980, Prog. of Theor. Phys., 64, 544.
\bibitem[Morbidelli et al. 2008]{mor08} Morbidelli, A., Crida, A., Masset, F., \& Nelson, R. P. 2008. A\&A, 478, 929
\bibitem[Mordasini et al. 2009a]{mord09a}Mordasini, C., Alibert, Y., \& Benz, W. 2009, A\&A, 501, 1139
\bibitem[Mordasini et al. 2009b]{mord09b}Mordasini, C., Alibert, Y., Benz, W., \& Naef, D. 2009, A\&A, 501, 1161
\bibitem[Nagasawa et al. 2003]{nag03}Nagasawa, M., Lin, D. N. C., \& Ida, S. 2003, ApJ, 586, 1374
\bibitem[Natta et al. 2006]{Nat06}Natta, A., Testi, L., \& Randich, S. 2006, A\&A, 452, 245
\bibitem[Nelson \& Papaloizou 2004]{NP04}Nelson, R. P., \& Papaloizou, J. C. B. 2004, MNRAS, 350, 849
\bibitem[Ogihara  \& Ida 2009]{OI09}Ogihara M., \& Ida, S. 2009, ApJ, 699, 824
\bibitem[Paardekooper \& Mellema]{PM06}Paardekooper, S.-J., \& Mellema, G. 2006, A\&A, 459, L17
\bibitem[Papaloizhou et al. 2001]{Pap01}Papaloizou, J. C. B., Nelson, R. P., \& Masset, F. 2001,  A\&A, 366, 263
\bibitem[Perri \& Cameron 1974]{PC74}Perri, F., \& Cameron, A. G. W. 1974, Icarus, 22, 416
\bibitem[Pollack et al 1996]{pollack}Pollack, J. B., Hubickyj, O., Bodenheimer, P., Lissauer, J. J., Podolak, M. \& Greenzweig, Y. 1996, Icarus, 124, 62
\bibitem[Pringle 1981]{pri81}Pringle, J. E. 1981, ARA\&A, 19, 137
\bibitem[Rasio \& Ford 1996]{RF96}Rasio, F. A., \& Ford, E. 1996, Science, 274, 954
\bibitem[Raymond et al. 2006]{Ray06}Raymond, S. N., Mandell, A. M., \& Sigurdsson, S. 2006, Science, 313, 1413
\bibitem[Ravikov 2006]{Rav06}Rafikov, R. R. 2006, ApJ, 648, 666
\bibitem[Safronov 1969]{Saf69}Safronov, V. S. 1969, Evolution of the Protoplanetary Cloud and Formation of the Earth and the planets, English translation NSSA TT F-677 (1972)
\bibitem[Sano et al. 2000]{sano00}Sano, T., Miyama, S. M., Umebayashi, T., \& Nakano, T. 2000, ApJ, 543, 486
\bibitem[Shakura \& Sunyaev 1973]{SS73}Shakura, N. I., \& Sunyaev, R. A. 1973, A\&A, 24, 337
\bibitem[Tanaka et al. (2002)]{Tan02}Tanaka, H., Takeuchi, T., \& Ward, W. R. 2002, ApJ, 565, 1257
\bibitem[Terquem \& Papaloizou(2007)]{TP07}Terquem, C., \&  Papaloizou, J. C. B. 2007, ApJ, 654, 1110
\bibitem[Thommes et al 2008]{Thommes}Thommes, E. W., Matsumura, S., \& Rasio, F. A. 2008, Science, 321, 814
\bibitem[Tsiganis et al. 2005]{tsi05}Tsiganis, K., Gomes, R., Morbidelli, A., \& Levison, H. F. 2005, Nature, 435, 459
\bibitem[Udry \& Santos (2007)]{ud07}Udry, S., \& Santos, N. C. 2007, Annu. Rev. Astro. Astrophys. 2007.45:397-439.
\bibitem[Vorobyov \& Basu 2009]{VB09}Vorobyov, E. I., \& Basu, I. 2009, ApJ, 703, 922
\bibitem[Wang \& Zhou 2010]{WZ10}Wang, S., \& Zhou, J. L., 2010, In preparation.
\bibitem[Ward 1988]{ward88}Ward, W. R. 1988, Icarus, 73, 330
\bibitem[Ward 1997]{Ward97}Ward, W. R. 1997, Icarus, 126, 261
\bibitem[Wittenmyer et al. 2009]{Wit09}Wittenmyer, R. A. et al. 2009, ApJS, 182, 97
\bibitem[Wright et al. 2009]{Wright}Wright, J. T., Upadhyay, S., Marcy, G. W., Fischer, D. A., Ford, E. B., \& Johnson, J. A. 2009, ApJ, 693, 1084
\bibitem[Yoshinaga et al. 1999]{Yosh09}Yoshinaga, K., Kokubo, E., \& Makino, J. 1999, Icarus, 139, 328
\bibitem[Zhou et al. 2005]{Zhou05} Zhou, J. L., Aarseth, S. J., Lin, D. N. C., \& Nagasawa, M. 2005, ApJ, 631, L85
\bibitem[Zhou et al. (2007)]{ZLS07}Zhou, J. L., Lin, D. N. C., \& Sun, Y. S. 2007, ApJ, 666, 423

\end{thebibliography}
\end{document}